\documentclass[12pt, draftclsnofoot, onecolumn]{IEEEtran}

\usepackage{amssymb}
\usepackage{amsmath}
\usepackage{cite}
\usepackage{url}
\usepackage{dsfont}
\usepackage{empheq}
\usepackage{xcolor}
\usepackage{graphicx}
\usepackage{subfigure}
\usepackage{enumitem}
\usepackage{fancyhdr}
\usepackage{mdwmath}	
\usepackage{mdwtab}
\usepackage{caption}
\usepackage{amsthm}
\usepackage{algorithm}
\usepackage{algorithmic}

\newtheorem{lemma}{Lemma}
\newtheorem{remark}{Remark}
\newtheorem{theorem}{Theorem}
\newtheorem{corollary}{Corollary}
\newtheorem{assumption}{Assumption}
\newtheorem{definition}{Definition}
\newtheorem{proposition}{Proposition}

\newcommand{\eqr}[1]{(\ref{#1})}
\newcommand{\fref}[1]{Fig.~\ref{#1}}

\newcommand*{\QEDA}{\null\nobreak\hfill\ensuremath{\blacksquare}}

\begin{document}
%\title{Energy-Efficient Federated Learning with Age-weighted FedSGD}
%\title{Age-weighted FedSGD for Device Selection in Federated Learning with Non-IID Data}
%\title{Age-weighted FedSGD for Device Selection in Federated Learning under Data Heterogeneity}
\title{Exploring Age-of-Information Weighting in Federated Learning under Data Heterogeneity}
\author{Kaidi~Wang,~\IEEEmembership{Member,~IEEE,}
Zhiguo~Ding,~\IEEEmembership{Fellow, IEEE,}
Daniel~K.~C.~So,~\IEEEmembership{Senior Member, IEEE,}
and Zhi~Ding,~\IEEEmembership{Fellow, IEEE}
\thanks{Kaidi Wang and Daniel K. C. So are with the Department of Electrical and Electronic Engineering, The University of Manchester, Manchester, M13 9PL, UK (email: kaidi.wang@ieee.org, d.so@manchester.ac.uk).}
\thanks{Zhiguo Ding is with Department of Electrical Engineering and Computer Science, Khalifa University, Abu Dhabi, UAE, and Department of Electrical and Electronic Engineering, University of Manchester, Manchester, UK (email: zhiguo.ding@manchester.ac.uk).}
\thanks{Zhi Ding is with the Department of Electrical and Computer Engineering, University of California at Davis, Davis, CA 95616 USA (email: zding@ucdavis.edu).}}
\maketitle
%%%%%%%%%%%%%%%%%%%%%%%%%%%%%%%%%%%%%%%%%%%%%%%%%
%%%%%%%%%%%%%%%%%%%%%%%%%%%%%%%%%%%%%%%%%%%%%%%%%
\begin{abstract}
This paper investigates federated learning in a wireless communication system, where random device selection is employed with non-independent and identically distributed (non-IID) data. The analysis indicates that while training deep learning networks using federated stochastic gradient descent (FedSGD) on non-IID datasets, device selection can generate gradient errors that accumulate, leading to potential weight divergence. To mitigate training divergence, we design an age-weighted FedSGD to scale local gradients according to the previous state of devices. To further improve learning performance by increasing device participation under the maximum time consumption constraint, we formulate an energy consumption minimization problem by including resource allocation and sub-channel assignment. By transforming the resource allocation problem into convex and utilizing KKT conditions, we derived the optimal resource allocation solution. Moreover, this paper develops a matching based algorithm to generate the enhanced sub-channel assignment. Simulation results indicate that i) age-weighted FedSGD is able to outperform conventional FedSGD in terms of convergence rate and achievable accuracy, and ii) the proposed resource allocation and sub-channel assignment strategies can significantly reduce energy consumption and improve learning performance by increasing the number of selected devices.
\end{abstract}
%%%%%%%%%%%%%%%%%%%%%%%%%%%%%%%%%%%%%%%%%%%%%%%%%
%%%%%%%%%%%%%%%%%%%%%%%%%%%%%%%%%%%%%%%%%%%%%%%%%
\section{Introduction}
With the spread of computer chips, powerful computational capabilities become available at edge nodes, and therefore, the collected data can be directly utilized for learning tasks \cite{chen2020flm}. In this context, federated learning, as a promising technology for distributed learning, has attracted considerable attention from both academia and industry. In federated learning, a neural network is constructed by a central server and shared among all participating devices \cite{mcmahan2017fl}. At each device, the received neural network is trained with local data and transmitted to the server for aggregation \cite{konevcny2016federated2}. Compared with centralized learning that requires offloading raw data to the server, in federated learning, learning tasks are executed collaboratively without data sharing, and hence, privacy security can be improved \cite{savazzi2021fl}. Furthermore, since the size of the transmitted neural network is generally smaller than the size of original data, communication efficiency can be achieved \cite{chen2021flm2}. However, since federated learning relies on periodic transmission, its performance is affected by wireless networks, and hence, the optimization of communications is recognized as an important research direction \cite{yang2022fl}.

Due to the fact that federated learning usually involves a large number of devices for multiple rounds of training and transmission, device selection/sampling becomes a common method to implement this algorithm under limited bandwidth resources \cite{nishio2019fl, fu2023fl}. Some existing works focused on addressing system heterogeneity by selecting devices based on hardware specifications and communication environment \cite{yang2020ds, chen2021fl1, guo2022fl, ribero2023fl}. In \cite{yang2020ds}, device selection and beamforming design were jointly considered in an over-the-air computation (AirComp) based federated learning framework, where an optimization problem was formulated to maximize the number of selected devices. By revealing the interaction between global loss and packet error rates, device selection was included to cope with the limited number of resource blocks \cite{chen2021fl1}. Particularly, in this work, a device can be selected only if the latency and energy consumption constraints can be satisfied. Since the transmitted models can be severely damaged by noise in AirComp based federated learning, in \cite{guo2022fl}, devices with weak channel conditions were ignored for aggregation as the transmit power is not sufficient to compensate for the effects of wireless communications. Recognizing the degradation of learning performance caused by low device availability, the authors of \cite{ribero2023fl} proposed a device selection strategy based on achievable long-term participation rates to mitigate the impact of device selection variance on global model convergence.

In realistic scenarios of federated learning, non-independent and identically distributed (non-IID) data is unevenly distributed among devices, which brings challenges to device selection \cite{konevcny2016federated, gafni2022flm}. Specifically, in system based device selection, the server tends to select devices with better channel conditions and/or powerful computational capacities, which may lead to a decline in learning performance on non-IID datasets \cite{zhao2018federated}. To this end, by selecting devices that provide more contributions in the aggregation, some works jointly considered system heterogeneity and data heterogeneity \cite{mohanmmad2021fl, luo2022fl, cho2022towards, kaidi2023fl2}. In \cite{mohanmmad2021fl}, channel conditions and local model updates were studied, and four device selection polices were proposed based on different priorities. Simulation results demonstrated that jointly including both metrics can provide better learning performance than using either metric separately. In order to achieve the target global loss within less time consumption, the selection probabilities of devices in the classic random device selection scheme were optimized based on latency and gradient norms \cite{luo2022fl}. Considering that the device contribution is not only related to dataset size, a biased device selection scheme was developed in \cite{cho2022towards}, in which the server transmits the global model to a set of candidate devices for evaluation, and then selects devices with larger local losses. In \cite{kaidi2023fl2}, age-of-information (AoI) was considered as a metric to improve the fairness of device selection. It was indicated that by minimizing the overall AoI of all devices, both learning performance and time consumption can be improved.

Since learning based device selection is performed on a set of available devices, its performance can be further improved by increasing device availability, which is determined by channel conditions, computational capacities, battery levels, etc. \cite{li2020fl, ribero2023fl}. Therefore, it is necessary to explore the optimal resource allocation based on these factors. Energy consumption, as an important criterion that can directly limit device participation, has been extensively researched in existing works \cite{yang2021fl, wang2023fl, alishahi2023fl, ren2023ecfl, lin2024cfl}. In \cite{yang2021fl}, a comprehensive energy consumption minimization problem was investigated in federated learning systems, where monotonicity analysis was utilized to obtain solutions. Wireless federated learning was also studied in eavesdropping scenarios, in which idle devices transmit jamming signals to improve the secrecy rate of the transmitting device \cite{wang2023fl}. In\cite{alishahi2023fl}, energy harvesting and non-orthogonal multiple access (NOMA) were exploited to provide computing energy and facilitate uplink transmission, respectively. In these works, the bisection method was utilized for algorithm design \cite{yang2021fl, wang2023fl, alishahi2023fl}. The authors of \cite{ren2023ecfl} focused on studying long-term energy consumption minimization, where deep reinforcement learning was employed. In \cite{lin2024cfl}, NOMA schemes were adopted in a clustered federated learning system, where sub-channel assignment and power allocation were studied to further enhance device availability.

As aforementioned, with non-IID data, system based device selection leads to a decline in learning performance \cite{yang2020ds, chen2021fl1, guo2022fl, ribero2023fl}, while learning based device selection requires additional transmission and analysis for local models or gradients \cite{mohanmmad2021fl, luo2022fl, cho2022towards}. A novel method, namely age-weighted FedSGD, is proposed to mitigate the learning performance degradation caused by implementing device selection on non-IID datasets. This scheme can be employed in a variety of existing device selection strategies without extra information transmission and model/data analysis, and hence, it will not increase system overhead or cause privacy leakage. Compared to \cite{kaidi2023fl2} which exploits AoI to guide device selection, in this work, the number of idle communication rounds serves as a weighting factor to directly regulate the update of global models. Moreover, to further increase device availability, an energy consumption minimization problem is jointly addressed through a low-complexity solution, thus avoiding the loss of optimality in previous works \cite{yang2021fl, wang2023fl, alishahi2023fl}. The main contributions can be summarized as follows:
\begin{itemize}[leftmargin=*]
\item A wireless federated learning network with random device selection is investigated. It is proved that in conventional federated stochastic gradient descent (FedSGD), device selection with non-IID data results in an error in global gradients, which is accumulated and amplified during training, thereby increasing weight divergence.
\item Based on the analyzed result, AoI is introduced to design age-weighted FedSGD, which can adjust the proportion of local gradients from selected devices in the global gradient. It is indicated that the proposed scheme enables the data distribution of the selected devices to approximate the global data distribution, thereby effectively reducing weight divergence.
\item In order to further improve the performance of device selection, an energy consumption minimization problem is formulated to increase device availability. By decoupling the formulated problem into two sub-problems, KKT conditions and matching theory are utilized to develop the closed-form resource allocation solution and sub-channel assignment algorithm, respectively.
\item Simulation results show that the proposed age-weighted FedSGD can significantly improve the performance of federated learning in the considered system, including convergence rate and achievable test accuracy. Moreover, KKT based resource allocation and matching based sub-channel assignment are able to minimize energy consumption and increase the number of selected devices.
\end{itemize}
%%%%%%%%%%%%%%%%%%%%%%%%%%%%%%%%%%%%%%%%%%%%%%%%%
%%%%%%%%%%%%%%%%%%%%%%%%%%%%%%%%%%%%%%%%%%%%%%%%%
\section{System Model}
Consider a wireless communication scenario for non-IID federated learning, where a server and $N$ devices collaborate to execute a learning task through $K$ sub-channels. All nodes are equipped with single-antennas. The collections of devices and sub-channels are represented by $\mathcal{N}=\{1,2,\cdots, N\}$ and $\mathcal{K}=\{1,2,\cdots, K\}$, respectively. It is assumed that the number of available sub-channels is less than the number of devices, and thus a subset of devices is randomly selected\footnote{Note that although this work considers classic random device selection, the proposed method can be utilized with multiple existing advanced device selection strategies.} to participate in the aggregation in each communication round, denoted by $\mathcal{S}_t$, where $|\mathcal{S}_t|\le K < N$. In each round, the global model and the device selection decision are transmitted from the server to all devices, and then the selected devices train the received global model using all local samples, as shown in the following:
\begin{equation}
f_n(\mathbf{w}^\mathrm{(t)}) =\frac{1}{\beta_n}\sum_{i=1}^{\beta_n} \ell(\mathbf{w}^\mathrm{(t)}; \boldsymbol{x}_{n,i}, y_{n,i}),
\end{equation}
where $\beta_n$ is the number of local samples at device $n$, $\mathbf{w}^\mathrm{(t)}$ is the global model in round $t$, $(\boldsymbol{x}_{n,i}, y_{n,i})$ is the $i$-th sample at device $n$. Given the set of selected devices $\mathcal{S}_t$, the global loss can be expressed as follows:
\begin{equation}
F(\mathbf{w}^\mathrm{(t)},\mathcal{S}_t)=\frac{\sum_{n\in\mathcal{S}_t}\beta_n f_n(\mathbf{w}^\mathrm{(t)})}{\sum_{n\in\mathcal{S}_t}\beta_n}.
\end{equation}
For aggregation, FedSGD is employed, where selected devices transmit local gradients to the server, and the global model is updated at the server, as follows:
\begin{equation}\label{updatefed}\vspace{-2mm}
\mathbf{w}^\mathrm{(t+1)}=\mathbf{w}^\mathrm{(t)}-\lambda \nabla F(\mathbf{w}^\mathrm{(t)},\mathcal{S}_t),
\end{equation}
where $\lambda$ is the learning rate.

In this paper, the following assumptions are considered.
\begin{assumption}\label{assume1}
With respect to $\mathbf{w}$, $\nabla F(\mathbf{w},\mathcal{N})$ is $L$-Lipschitz continuous, i.e., 
\begin{equation}
\|\nabla F(\mathbf{w}^{\mathrm{(t-1)}}, \mathcal{N})\!-\!\nabla F(\mathbf{w}^\mathrm{(t)}, \mathcal{N})\|\le L\|\mathbf{w}^{\mathrm{(t-1)}}\!-\!\mathbf{w}^\mathrm{(t)}\|.
\end{equation}
\end{assumption}
\begin{assumption}
The global loss function $F(\mathbf{w}^\mathrm{(t)}, \mathcal{N})$ satisfies the Polyak-Lojasiewicz inequality with positive parameter $\mu$, as shown in follows:
\begin{equation}
\|\nabla F(\mathbf{w}^\mathrm{(t)}, \mathcal{N})\|^2\ge 2\mu \left[F(\mathbf{w}^\mathrm{(t)}, \mathcal{N})-F(\mathbf{w}^{*}, \mathcal{N})\right].
\end{equation}
\end{assumption}
%%%%%%%%%%%%%%%%%%%%%%%%%%%%%%%%%%%%%%%%%%%%%%%%%
\subsection{Weight Divergence in Conventional FedSGD}
Since device selection is implemented in a federated learning algorithm using non-IID datasets, the data distribution of the selected devices may be different from the global data distribution, and therefore, the weight divergence issue may occur \cite{zhao2018federated, kaidi2023fl2}. That is, the divergence between the weights obtained from the considered federated learning framework and the desired weights obtained from centralized learning increases with training. To evaluate weight divergence, complete device selection is included as the baseline, where all devices are selected in each communication round. The updated of the true global model in this case is given by
\begin{equation}\label{updatetrue}
\mathbf{w}_\mathrm{T}^\mathrm{(t+1)} = \mathbf{w}_\mathrm{T}^\mathrm{(t)}-\lambda \nabla F(\mathbf{w}_\mathrm{T}^\mathrm{(t)}, \mathcal{N}),
\end{equation}
where
\begin{equation}
F(\mathbf{w}_\mathrm{T}^\mathrm{(t)}, \mathcal{N}) = \frac{\sum_{n\in\mathcal{N}}\beta_n f_n(\mathbf{w}_\mathrm{T}^\mathrm{(t)})}{\sum_{n\in\mathcal{N}}\beta_n}.
\end{equation}
Note that the complete device selection scheme can be treated as centralized learning, since in the considered federated learning algorithm, all local data is utilized for training and the number of local epochs is one. Based on the definition of the true global model, the following theorem can be obtained.
\begin{theorem}\label{weightdivergence}
Defining the error caused by device selection as the difference in the global loss gradient between random device selection and complete device selection, i.e.,
\begin{equation}\label{error}
\mathbf{e}^\mathrm{(t)}\triangleq\nabla F(\mathbf{w}^\mathrm{(t)}, \mathcal{S}_t)\!-\!\nabla F(\mathbf{w}^\mathrm{(t)}, \mathcal{N}),
\end{equation}
the weight divergence in the considered federated learning framework is bounded by:
\begin{equation}
\left\|\mathbf{w}^{\mathrm{(t+1)}}-\mathbf{w}_\mathrm{T}^{\mathrm{(t+1)}}\right\|\le (1\!+\!\lambda L)^t\left\|\mathbf{w}^{\mathrm{(1)}}\!-\!\mathbf{w}_\mathrm{T}^{\mathrm{(1)}}\right\|\!+\!\lambda\left\|\sum_{i=1}^{t}\mathbf{e}^{(i)}\right\|\!+\!\lambda^2L\!\sum_{j=1}^{t-1}(1\!+\!\lambda L)^{j-1}\!\left\|\sum_{i=1}^{t-j}\!\mathbf{e}^{(i)}\right\|.
\label{wdexpression}
\end{equation}
\begin{IEEEproof}
Refer to Appendix~A.
\end{IEEEproof}
\end{theorem}

Theorem \ref{weightdivergence} indicates that in the considered federated learning framework, device selection leads to weight divergence, and this effect can be described as the error of the global loss gradient in each communication round. Based on Theorem \ref{weightdivergence}, the following remarks can be obtained.
\begin{remark}
In the considered federated learning algorithm, the weight divergence is mainly caused by two parts, including the difference between initial global models, i.e., $\|\mathbf{w}^{\mathrm{(1)}}-\mathbf{w}_\mathrm{T}^{\mathrm{(1)}}\|$, and the accumulated error, i.e., $\|\sum_i\mathbf{e}^{(i)}\|$.
\end{remark}
\begin{remark}
The impact of the accumulated error from previous rounds is amplified with training, since $1+\lambda L>1$. That is, the impact of errors in the early stages of training plays a major role in weight divergence.
\end{remark}
\begin{remark}
When utilizing different initial global models, even if complete device selection is applied, i.e., the error is zero, large weight divergence may still be encountered.
\end{remark}

It can be observed from Theorem \ref{weightdivergence} that the weight divergence can be mitigated by reducing $\|\sum_{i=1}^t\mathbf{e}^{(i)}\|$. To this end, conventional methods, such as importance sampling, focus on reducing $\|\mathbf{e}^\mathrm{(t)}\|$ in each communication round by bringing $\nabla F(\mathbf{w}^\mathrm{(t)}, \mathcal{S}_t)$ closer to $\nabla F(\mathbf{w}^\mathrm{(t)}, \mathcal{N})$ \cite{fu2023fl, rizk2022fl}. However, these methods need to analyze local gradients, which is impractical in the realistic scenario due to the high-complexity or privacy issues. Inspired by the fact that the errors are accumulated, this paper introduces a weighting factor\footnote{In this paper, the terms “weighting factor” and “weights” refer to the local gradient adjustment coefficient and neural network parameters, respectively.} to scale the error of the current round according to the accumulated error from previous rounds, thereby reducing weight divergence.  Specifically, by treating $\mathbf{e}^\mathrm{(t)}$ and $\sum_{i=1}^{t-1}\mathbf{e}^{(i)}$ as two vectors, if the elements in $\mathbf{e}^\mathrm{(t)}$ are close to the corresponding elements in $\sum_{i=1}^{t-1}\mathbf{e}^{(i)}$, a small weighting factor is adopted to reduce these elements; otherwise, these elements are amplified. As a result, the accumulated error $\sum_{i=1}^{t-1}\mathbf{e}^{(i)}$ can be compensated by using $\mathbf{e}^\mathrm{(t)}$, and the value of $\|\sum_{i=1}^t\mathbf{e}^{(i)}\|$ can be reduced.
%%%%%%%%%%%%%%%%%%%%%%%%%%%%%%%%%%%%%%%%%%%%%%%%%
\subsection{Age-weighted FedSGD}
Based on the definition of error in \eqr{error}, weighting factor should be applied to $\nabla F(\mathbf{w}^\mathrm{(t)}, \mathcal{S}_t)$, since $\nabla F(\mathbf{w}^\mathrm{(t)}, \mathcal{N})$ is unknown in practical training. In other words, the weighting factor should be designed according to the difference in device selection between communication rounds. In particular, in the considered system, some devices may need to wait several rounds before participating in the aggregation. In this case, AoI is introduced to record recent device selection status and generate weighting factors \cite{yang2020aoi, kaidi2022fl}. For device $n$, its AoI in round $t$ is defined as follows:
\begin{equation}
A_n^\mathrm{(t)}=\left\{\begin{array}{ll}
1, &\quad\text{if}\quad n\in\mathcal{S}_{t-1},\\
A_n^\mathrm{(t-1)}+1, &\quad\text{if}\quad n\notin\mathcal{S}_{t-1}.
\end{array}\right.
\end{equation}
The above equation indicates that if device $n$ is selected in last round, its AoI becomes $1$; otherwise, it increases by $1$. Based on this definition, the age-weighted FedSGD is proposed with the following weighting factor\footnote{The AoI based weighting factor presented in \eqr{aoiweight} is for simplicity. It is observed that learning performance is highly sensitive to AoI, and other expressions of the weighting factor may provide further improvements.}:
\begin{equation}\label{aoiweight}
\omega_n^\mathrm{(t)}=\frac{A_n^\mathrm{(t)}|\mathcal{S}_t|}{\sum_{i\in\mathcal{S}_t}A_i^\mathrm{(t)}},
\end{equation}
where $|\mathcal{S}_t|$ is included for normalization. At the server, the global model is updated based on the age-weighted local gradients, as follows:
\begin{equation}\label{update}
\mathbf{w}^\mathrm{(t+1)}=\mathbf{w}^\mathrm{(t)}-\lambda\nabla G(\mathbf{w}^\mathrm{(t)},\mathcal{S}_t),
\end{equation}
where
\begin{equation}
G(\mathbf{w}^\mathrm{(t)},\mathcal{S}_t)=\frac{\sum_{n\in\mathcal{S}_t}\omega_n^\mathrm{(t)}\beta_nf_n(\mathbf{w}^\mathrm{(t)})}{\sum_{n\in\mathcal{S}_t}\beta_n}.
\end{equation}
Note that the AoI of all devices can be counted at the server, and thus the proposed scheme does not require additional information transmission. Moreover, by including the AoI based weighting factor in the update of local models, the proposed approach can also be utilized in federated averaging (FedAvg), where the AoI of all devices can be transmitted together with the global model.

\begin{figure}[!t]
\centering{\includegraphics[width=100mm]{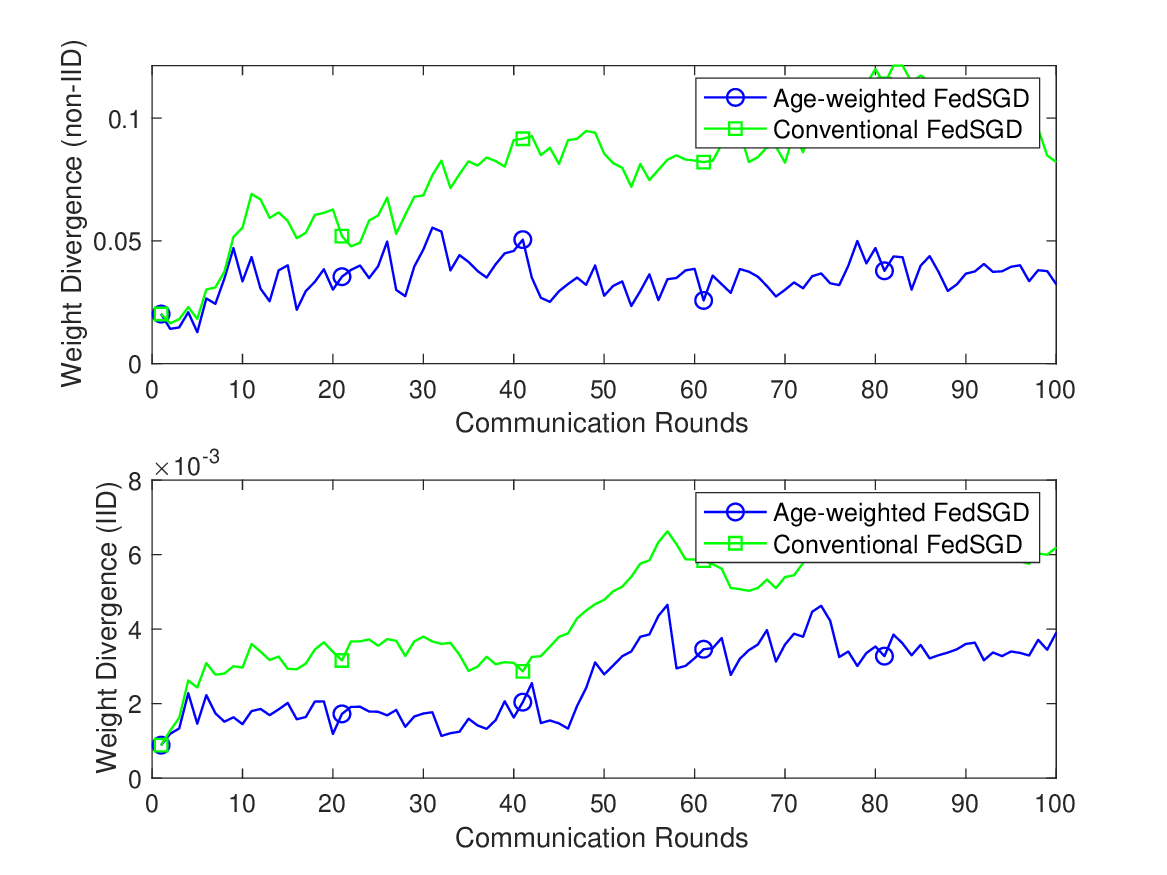}}
\caption{An empirical result to validate the performance of age-weighted FedSGD on MNIST dataset. $N=10$ and $K=5$.}
\label{wdfigure2}
\vspace{-4mm}
\end{figure}

The performance of age-weighted FedSGD and conventional FedSGD is compared in \fref{wdfigure2}, where weight divergence is calculated by $\|\mathbf{w}^\mathrm{(t)}-\mathbf{w}_\mathrm{T}^\mathrm{(t)}\|$. It can be observed that on non-IID datasets, weight divergence in conventional FedSGD increases with training, while age-weighted FedSGD can efficiently reduce weight divergence and control it to a certain level. Furthermore, it is worth pointing out that age-weighted FedSGD is still valid with IID data, although the weight divergence issue is not severe in this case.
%%%%%%%%%%%%%%%%%%%%%%%%%%%%%%%%%%%%%%%%%%%%%%%%%
\subsection{Performance Analysis}
In this subsection, age-weighted FedSGD is analyzed for multi-class classification. Consider a $C$-class classification problem with compact space $\mathcal{X}$ and label space $\mathcal{Y}=\mathcal{C}$, where $\mathcal{C}=\{1,2,\dots, C\}$. The data distribution of device $n$ is defined as follows:
\begin{equation}
P_n=\left[\frac{\sum_{i=1}^{\beta_n}\!\mathds{1}_{y_{n,i}=c}}{\beta_n}\bigg|\forall c\in\mathcal{C}\right],
\end{equation}
By adopting the cross-entropy loss, the local loss is given by
 \begin{align}\nonumber
f_n(\mathbf{w}^\mathrm{(t)})&=\mathbb{E}_{\mathbf{x}, y\sim P_n}\!\left[\sum_{c\in\mathcal{C}}\mathds{1}_{y=c}\log f_c(\boldsymbol{x},\mathbf{w}^\mathrm{(t)})\right]\\
&=\sum_{c\in\mathcal{C}}P_n(y=c)\mathbb{E}_{\boldsymbol{x}|y=c}\!\left[\log f_c(\boldsymbol{x},\mathbf{w}^\mathrm{(t)})\right],
 \end{align}
where $f_c(\boldsymbol{x},\mathbf{w}^\mathrm{(t)})$ indicates the probability for class $c$, and the sample $(\boldsymbol{x}, y)$ follows data distribution $P_n$. In this case, by defining $P_{\mathcal{S}_t}$ and $P_{\mathcal{N}}$ as the data distributions of the selected devices and all devices, i.e., 
\begin{align}
P_{\mathcal{S}_t}\!=\!\left[\frac{\sum_{n\in\mathcal{S}_t}\!\sum_{i=1}^{\beta_n}\!\mathds{1}_{y_{n,i}=c}}{\sum_{n\in\mathcal{S}_t}\!\beta_n}\bigg|\forall c\in\mathcal{C}\right]\!=\!\frac{\sum_{n\in\mathcal{S}_t}\!\beta_n P_n}{\sum_{n\in\mathcal{S}_t}\!\beta_n},
\end{align}
and
\begin{equation}
P_\mathcal{N}\!=\!\left[\frac{\sum_{n\in\mathcal{N}}\!\sum_{i=1}^{\beta_n}\!\mathds{1}_{y_{n,i}=c}}{\sum_{n\in\mathcal{N}}\!\beta_n}\bigg|\forall c\in\mathcal{C}\right]\!=\!\frac{\sum_{n\in\mathcal{N}}\!\beta_n P_n}{\sum_{n\in\mathcal{N}}\!\beta_n},
\end{equation}
the error in \eqr{error} can be expressed as
\begin{equation}\label{errormulticlass}
\mathbf{e}^\mathrm{(t)}\!=\!\sum_{c\in\mathcal{C}}[P_{\mathcal{S}_t}\!(y\!=\!c)\!-\!\!P_{\mathcal{N}}(y\!=\!c)]\nabla\mathbb{E}_{\boldsymbol{x}|y=c}\!\left[\log f_c(\boldsymbol{x},\mathbf{w}^\mathrm{(t)})\right].
\end{equation}
Based on this equation, the following remarks can be obtained.
\begin{remark}
In the multi-class classification problem, the error caused by device selection is mainly determined by the distance between the data distribution of the selected devices $P_{\mathcal{S}_t}$ and the global data distribution $P_{\mathcal{N}}$, and this impact is affected by the gradient $\nabla\mathbb{E}_{\boldsymbol{x}|y=c}\left[\log f_c(\boldsymbol{x},\mathbf{w}^\mathrm{(t)})\right]$.
\end{remark}
\begin{remark}
In the considered federated learning framework, if the data distribution of the selected devices is the same as the global data distribution, the weight divergence issue can be avoided.
\end{remark}

\begin{figure}[!t]
\centering{\includegraphics[width=120mm]{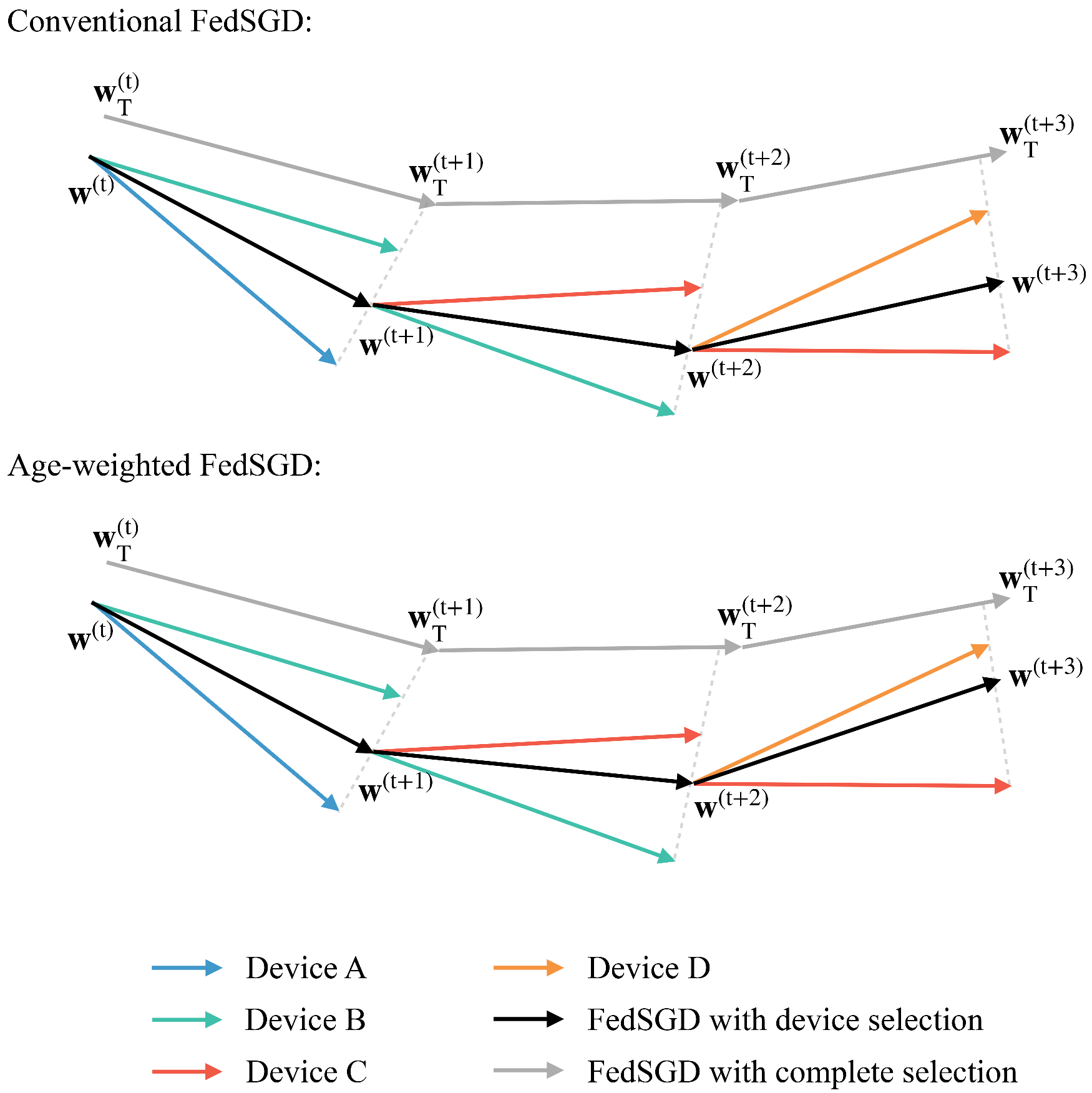}}
\caption{An illustration of weight divergence for federated learning with conventional FedSGD and age-weighted FedSGD.}
\label{wdfigure3}
\end{figure}

According to Theorem \ref{weightdivergence}, the impact of the error in any round is amplified with the increasing of communication rounds. That is, in the multi-class classification problem, the impact of data distributions caused by device selection is different due to the sequence of selection. Moreover, the data distribution $P_{\mathcal{S}_t}$ is determined by set $\mathcal{S}_t$, and the devices of this set may have different selection records. However, in conventional FedSGD, this difference is not reflected. In the proposed age-weighted FedSGD, the data distribution of set $\mathcal{S}_t$ is given by
\begin{equation}
P_{\mathcal{S}_t}^\omega\!=\!\left[\frac{\sum_{n\in\mathcal{S}_t}\!\omega_n^\mathrm{(t)}\!\sum_{i=1}^{\beta_n}\!\mathds{1}_{y_{n,i}=c}}{\sum_{n\in\mathcal{S}_t}\!\beta_n}\bigg|\forall c\in\mathcal{C}\right]\!=\!\frac{\sum_{n\in\mathcal{S}_t}\!\omega_n^\mathrm{(t)}\!\beta_n P_n}{\sum_{n\in\mathcal{S}_t}\!\beta_n}.
\end{equation}
It can be seen that except data size $\beta_n$, another weighting factor $\omega_n^\mathrm{(t)}$ is included to  scale the influence from different devices based on the previous device selection results. Specifically, since the impact of $\mathcal{S}_{t-1}$ is amplified in round $t$, a big weighting factor should be included to the devices in $\mathcal{N}\backslash\mathcal{S}_{t-1}$. Due to the same reason, when any device in $\mathcal{S}_{t-1}$ is selected in round $t$, a small weighting factor should be included. \fref{wdfigure3} is included to explain age-weighted FedSGD, where $2$ devices are selected in each communication round. Compared to conventional FedSGD, the global model obtained in round $t+2$, i.e., $\mathbf{w}^\mathrm{(t+2)}$, is closer to the local model of device C, since the gradients used in model updates are weighted according to the device selection result of round $t+1$. Due to the same reason, device D dominates the aggregation in round $t+3$. As a result, the distance between $\mathbf{w}^\mathrm{(t+3)}$ and true global model $\mathbf{w}_\mathrm{T}^\mathrm{(t+2)}$ is reduced by utilizing age-weighted FedSGD.

With age-weighted FedSGD, the difference in global gradient between random device selection and complete device selection is given by
\begin{equation}\label{cond1}
\mathbf{g}^\mathrm{(t)}\triangleq\nabla G(\mathbf{w}^\mathrm{(t)},\mathcal{S}_t)-\nabla F(\mathbf{w}^\mathrm{(t)}, \mathcal{N}).
\end{equation}
Since all devices are selected in complete device selection, $A_n^\mathrm{(t)}=1, \forall n, t$ always holds, and hence, the AoI based weighting factor satisfies $\omega_n^\mathrm{(t)}=1$. In this case, the following equation can be obtained:
\begin{equation}\label{cond2}
\nabla G(\mathbf{w}^\mathrm{(t)}, \mathcal{N})=\nabla F(\mathbf{w}^\mathrm{(t)}, \mathcal{N}).
\end{equation}
The above equation indicates that age-weighted FedSGD can be utilized for complete device selection without any impact. Based on \eqr{cond1} and \eqr{cond2}, the expected convergence rate of age-weighted FedSGD can be obtained.
\begin{theorem}\label{convergence}
With age-weighted FedSGD, the expected reduction of global loss in round $t$ is bounded by
\begin{equation}
\mathbb{E}\!\left[F(\mathbf{w}^\mathrm{(t+1)}, \mathcal{N})\!-\!F(\mathbf{w}^{*})\right]\!\le \!\left(\!1\!\!-\!\!\frac{\mu}{L}\!\right)^t\!\!\mathbb{E}\!\left[\!F(\mathbf{w}^{\mathrm{(1)}}, \mathcal{N})\!-\!\!F(\mathbf{w}^{*})\!\right]\!\!+\!\!\frac{1}{2L}\!\sum_{i=1}^{t}\!\left(\!1\!\!-\!\!\frac{\mu}{L}\!\right)^{t-i}\!\!\mathbb{E}\!\left[\!\|\mathbf{g}^{(i)}\|^2\!\right],\!\!\!
\end{equation}
where
\begin{equation}
\mathbb{E}\!\left[\|\mathbf{g}^\mathrm{(t)}\|^2\right]=\left(\!1\!-\!\frac{|\mathcal{S}_t|}{N}\right)\!\!\frac{\sum_{n\in\mathcal{N}}\!\beta_n^2\left\|\omega_n^\mathrm{(t)}\nabla f_n(\mathbf{w}^\mathrm{(t)})\!-\!\!\nabla F(\mathbf{w}^\mathrm{(t)}, \mathcal{N})\right\|^2}{|\mathcal{S}_t|(N\!-\!1)(\frac{1}{N}\!\sum_{n\in\mathcal{N}}\!\beta_n)^2}.
\end{equation}
\begin{IEEEproof}
Refer to Appendix~B.
\end{IEEEproof}
 \end{theorem}
Theorem \ref{convergence} indicates that the convergence rate of the considered federated learning algorithm can be improved by increasing the number of selected devices, i.e., $|\mathcal{S}_t|$. On the other hand, the weight divergence issue will be more severe if some devices are unavailable, since the data distribution in this case is different from the global data distribution. As a result, in order to improve device availability,  the optimization of the computation and transmission stages needs to be considered.
%%%%%%%%%%%%%%%%%%%%%%%%%%%%%%%%%%%%%%%%%%%%%%%%%
\subsection{Local Training and Transmission}
For any selected device $n$ assigned to sub-channel $k$, it trains the global model based on all local samples, and hence, the computing time can be expressed as follows:
\begin{equation}
T_{k,n}^{\mathrm{cp}} = \frac{\mu\beta_n}{\tau_{k,n} C_n},
\end{equation}
where $\mu$ is the required number of cycles to train each sample, $\tau_{k,n}$ is the computing resource allocation coefficient, and $C_n$ is the computational capacity of device $n$. According to \cite{yang2021fl, kaidi2023fl2}, the corresponding energy consumption for local training is given by
\begin{equation}
E_{k,n}^{\mathrm{cp}} = \kappa\mu\beta_n(\tau_{k,n} C_n)^2,
\end{equation}
where $\kappa$ is the power consumption coefficient of each central processing unit (CPU) cycle. After local training, the local gradient is sent to the server through the assigned sub-channel at the following data rate: 
\begin{equation}\label{datarate}
R_{k,n}=B\log_2(1+\alpha_{k,n}P_n|h_{k,n}|^2),
\end{equation}
where $B$ is the allocated bandwidth of each sub-channel, $\alpha_{k,n}$ is the power allocation coefficient, $P_n$ is the maximum transmit power, $|h_{k,n}|^2=\eta|g_n|^2 d_n^{-a}\sigma^{-2}$ is the normalized channel gain, $\eta$ is the frequency dependent factor, $g_n$ is the small-scale fading coefficient, $d_n$ is the distance between device $n$ and the server, $\alpha$ is the path loss exponent, and $\sigma^{2}$ is the noise power. The communication time of device $n$ assigned to sub-channel $k$ can be expressed as follows:
\begin{equation}
T_{k,n}^{\mathrm{cm}}=\frac{D}{R_{k,n}},
\end{equation}
where $D$ is the size of the local gradient for each device. The energy consumption for transmission is given by
\begin{equation}
E_{k,n}^{\mathrm{cm}} = \alpha_{k,n}P_nT_{k,n}^{\mathrm{cm}}.
\end{equation}

%%%%%%%%%%%%%%%%%%%%%%%%%%%%%%%%%%%%%%%%%%%%%%%%%
%%%%%%%%%%%%%%%%%%%%%%%%%%%%%%%%%%%%%%%%%%%%%%%%%
\section{Problem Formulation}
In order to improve learning performance by increasing device availability, an energy consumption minimization problem is formulated under the maximum time consumption constraint, as shown in follows:
\begin{subequations}
\begin{empheq}{align}
\min_{\boldsymbol{\tau}, \boldsymbol{\alpha}, \boldsymbol{\psi}}\quad & \sum_{n\in\mathcal{S}_t}\sum_{k\in\mathcal{K}}\psi_{k,n}(E_{k,n}^{\mathrm{cp}}+E_{k,n}^{\mathrm{cm}})\\
\textrm{s.t.} \quad & T_{k,n}^{\mathrm{cp}}+T_{k,n}^{\mathrm{cm}}\le T_{n}^{\mathrm{max}}, \forall k \in\mathcal{K}, \forall n \in\mathcal{S}_t,\\
& \tau_{k,n} \in [0, 1], \forall k \in\mathcal{K}, \forall n \in\mathcal{S}_t, \\
& \alpha_{k,n} \in[0, 1], \forall k \in\mathcal{K}, \forall n \in\mathcal{S}_t, \\
& \psi_{k,n}^\mathrm{(t)}\in\{0, 1\},  \forall k \in\mathcal{K},  \forall n \in\mathcal{S}_t,\\
& \sum\nolimits_{n\in\mathcal{S}_t}\psi_{k,n}^\mathrm{(t)} \in \{0, 1\}, \forall k \in\mathcal{K},\\
& \sum\nolimits_{k\in\mathcal{K}} \psi_{k,n}^\mathrm{(t)} \in \{0, 1\}, \forall n \in\mathcal{S}_t,
\end{empheq}
\label{problem}
\end{subequations}\vspace{-2mm}\\
where $\boldsymbol{\tau}$, $\boldsymbol{\alpha}$, and $\boldsymbol{\psi}$ are the sets of all computing resource allocation coefficients, power allocation coefficients, and sub-channel assignment indicators, respectively. In constraint (\ref{problem}b), $T_{n}^{\mathrm{max}}$ denotes the maximum time consumption of each communication round. Constraints (\ref{problem}c) and (\ref{problem}d) indicate that computing resource allocation coefficients and power allocation coefficients range from $0$ to $1$. Constraints (\ref{problem}e), (\ref{problem}f) and (\ref{problem}g) represent that the sub-channel assignment indicator is a binary variable, any sub-channel can be occupied by at most one device, and any device can be assigned to at most one sub-channel, respectively. In particular, in problem \eqr{problem}, resource allocation and sub-channel assignment are performed with the given set of selected devices. 

Due to the fact that the formulated problem is a mixed integer linear programming problem, it is decoupled into two sub-problems and solved iteratively. With the fixed sub-channel assignment, the resource allocation problem can be presented as follows:
\begin{subequations}
\begin{empheq}{align}
\min_{\boldsymbol{\tau}, \boldsymbol{\alpha}}\quad & \sum_{n\in\mathcal{S}_t}\sum_{k\in\mathcal{K}}E_{k,n}^{\mathrm{cp}}+E_{k,n}^{\mathrm{cm}}\\\nonumber
\textrm{s.t.} \quad & \text{(\ref{problem}b), (\ref{problem}c), and (\ref{problem}d)}.
\end{empheq}
\label{raproblem}
\end{subequations}\vspace{-2mm}\\
By removing the constraints related to resource allocation, the sub-channel assignment problem is shown in follows:
\begin{subequations}
\begin{empheq}{align}
\min_{\boldsymbol{\psi}}\quad & \sum_{n\in\mathcal{S}_t}\sum_{k\in\mathcal{K}}\psi_{k,n}(E_{k,n}^{\mathrm{cp}}+E_{k,n}^{\mathrm{cm}})\\\nonumber
\textrm{s.t.} \quad & \text{(\ref{problem}e), (\ref{problem}f), and (\ref{problem}g)}.
\end{empheq}
\label{saproblem}
\end{subequations}

%%%%%%%%%%%%%%%%%%%%%%%%%%%%%%%%%%%%%%%%%%%%%%%%%
%%%%%%%%%%%%%%%%%%%%%%%%%%%%%%%%%%%%%%%%%%%%%%%%%
\section{Joint Optimization of Computational Resource Allocation and Power Allocation}
Since the adjustment of resource allocation coefficients for any device cannot affect other devices, the resource allocation problem in \eqr{raproblem} is divided into $K$ sub-problems and solved independently. The resource allocation problem for device $n$ assigned to sub-channel $k$ is given by
\vspace{-1mm}\begin{subequations}
\begin{empheq}{align}
\min_{\tau_{k,n}, \alpha_{k,n}}\quad\!\!\!\!\! & \kappa\mu\beta_n(\tau_{k,n} C_n)^2\!+\!\frac{\alpha_{k,n}P_nD}{B\log_2(1\!+\!\alpha_{k,n}P_n|h_{k,n}|^2)}\\
\textrm{s.t.} \quad & \!\frac{\mu\beta_n}{\tau_{k,n} C_n}\!+\!\frac{D}{B\log_2(1\!+\!\alpha_{k,n}P_n|h_{k,n}|^2)}\le T_{n}^{\mathrm{max}},\!\!\\
& \tau_{k,n} \in [0, 1],\\
& \alpha_{k,n} \in[0, 1].
\end{empheq}
\label{raproblem1}
\end{subequations}\vspace{-2mm}\\
Note that the above problem is infeasible if any constraint is not satisfied. Hence, the following remark can be drawn.
\vspace{-1mm}\begin{remark}
For any device $n$ assigned to sub-channel $k$, its local gradient cannot be transmitted if
\vspace{-1mm}\begin{equation}
\frac{\mu \beta_n}{C_n}+\frac{D}{B\log_2(1+P_n|h_{k,n}|^2)}>T_{n}^{\mathrm{max}}.
\end{equation}
\end{remark}
That is, the selected devices may not be able to transmit local gradients to the server with the given time limitation, even if all resources are utilized. On the other hand, if this condition does not hold, the training and transmission tasks can alway be completed, which means $\tau_{k,n}>0$ and $\alpha_{k,n} > 0$. Therefore, $x_1$ and $x_2$ are introduced to replace the optimization variables, where $x_1=1/\tau_{k,n}$ and $x_2=1/[B\log_2(1+\alpha_{k,n}P_n|h_{k,n}|^2)]$. Problem \eqr{raproblem1} can be equivalently transformed as follows:
\vspace{-1mm}\begin{subequations}
\begin{empheq}{align}
\min_{\mathbf{x}}\quad &\frac{\kappa\mu\beta_nC_n^2}{x_1^2}\!+\!x_2\frac{(2^\frac{1}{x_2B}-1)D}{|h_{k,n}|^2},\\
\textrm{s.t.} \quad & \mu \beta_n C_n^{-1}x_1+Dx_2\le T_{n}^{\mathrm{max}},\\
& x_1 \ge 1,\\
& x_2 \ge \frac{1}{B\log_2(1+P_n|h_{k,n}|^2)},
\end{empheq}
\label{raproblem3}
\end{subequations}\vspace{-2mm}\\
where $\mathbf{x}=\{x_1, x_2\}$. It can be proved that the above problem is convex and satisfies Slater's condition, and hence, KKT conditions are utilized to derive the optimal solution \cite{boyd2004convex}. By introducing the Lagrangian multiplier $\lambda_i$ for the inequality constraints, the Lagrangian function is given by
\begin{align}\nonumber
\!L(\mathbf{x}) &\!=\! \frac{\kappa\mu\beta_nC_n^2}{x_1^2}\!+\!x_2\frac{(2^\frac{1}{x_2B}\!-\!1)D}{|h_{k,n}|^2}\!+\!\lambda_1\!\!\left(\!\frac{\mu \beta_n}{C_n}x_1\!\!+\!\!Dx_2\!\!-\!T_{n}^{\mathrm{max}}\!\!\right)\\
&\quad\!+\!\lambda_2(1\!-\!x_1)\!+\!\lambda_3\!\!\left[\frac{1}{B\log_2(1\!\!+\!\!P_n|h_{k,n}|^2)}\!-\!x_2\right].
\label{lagrangian}
\end{align}
Based on the Lagrangian function, the optimal solution of problem \eqr{raproblem3} can be presented below.
\vspace{-1mm}\begin{proposition}\label{closedform}
In case of $\mu \beta_n C_n^{-1}+D\upsilon_1\le T_{n}^{\mathrm{max}}$, by defining
\vspace{-2mm}\begin{equation}\label{vdefine}
\left\{\begin{array}{ll}
&\upsilon_1\triangleq \dfrac{1}{B\log_2(1+P_n|h_{k,n}|^2)},\\
&\upsilon_2\triangleq \dfrac{1}{B(T_{n}^{\mathrm{max}}-\mu\beta_nC_n^{-1})},
\end{array}\right.
\end{equation}
the optimal solution of problem \eqr{raproblem3} is given by
\begin{enumerate}[leftmargin=*]
\item $x_1^*=1$ and $x_2^*=\upsilon_1$, if the following condition holds:
\begin{equation}
\mu \beta_n C_n^{-1}+D\upsilon_1=T_{n}^{\mathrm{max}}.
\end{equation}
\item  $x_1^*=1$ and $x_2^*=(T_{n}^{\mathrm{max}}-\mu\beta_nC_n^{-1})/D$, if
\begin{equation}
\left\{\begin{array}{ll}
&\!\!\!\!\!\!\!\!\mu \beta_n C_n^{-1}+D\upsilon_1<T_{n}^{\mathrm{max}},\\
&\!\!\!\!\!\!\!\!D\upsilon_2\ln(2)2^{D\upsilon_2}\!\!-\!2^{D\upsilon_2}\!\!+\!1\!-\!2\kappa C_n^3|h_{k,n}|^2>0.\!\!\!\!\!\!\!\!
\end{array}\right.
\end{equation}
\item $x_1^*=(T_{n}^{\mathrm{max}}-D\upsilon_1)C_n(\mu\beta_n)^{-1}$ and $x_2^*=\upsilon_1$ if 
\begin{equation}
\left\{\begin{array}{ll}
&\!\!\!\!\!\!\!\!\!\mu \beta_n C_n^{-1}+D\upsilon_1<T_{n}^{\mathrm{max}},\\
&\!\!\!\!\!\!\!\!\!2^\frac{1}{B\upsilon_1}\!-\!1\!-\!\dfrac{1}{B\upsilon_1}\!\ln(2)2^\frac{1}{B\upsilon_1}\!\!+\!\dfrac{2\kappa(\mu\beta_n)^3|h_{k,n}|^2}{(T_{n}^{\mathrm{max}}\!-\!D\upsilon_1)^3}>0.\!\!\!\!\!\!\!\!\!\!
\end{array}\right.
\end{equation}
\item Otherwise, the optimal solution can be obtained by solving the following equations:
\vspace{-2mm}\begin{equation}
\left\{\begin{array}{ll}
&\dfrac{2\kappa C_n^3}{(x_1^*)^3}= \dfrac{\ln(2)2^\frac{1}{Bx_2^*}}{B|h_{k,n}|^2x_2^*}-\dfrac{2^\frac{1}{Bx_2^*}-1}{|h_{k,n}|^2},\vspace{1mm}\\
&\mu \beta_n C_n^{-1}x_1^*+Dx_2^*-T_{n}^{\mathrm{max}}=0.
\end{array}\right.
\end{equation}
\end{enumerate}
\begin{IEEEproof}
Refer to Appendix~C.
\end{IEEEproof}
\end{proposition}
According to the above proposition, the optimal solution of problem \eqr{raproblem1} can be obtained as follows:
\begin{equation}
\left\{\begin{array}{ll}
&\tau_{k,n}^*=1/x_1^*,\\
&\alpha_{k,n}^*=\dfrac{2^\frac{1}{Bx_2^*}-1}{P_n|h_{k,n}|^2},
\end{array}\right.
\end{equation}
and the formulated problem in \eqr{raproblem} is jointly solved.
%%%%%%%%%%%%%%%%%%%%%%%%%%%%%%%%%%%%%%%%%%%%%%%%%
%%%%%%%%%%%%%%%%%%%%%%%%%%%%%%%%%%%%%%%%%%%%%%%%%
\section{Matching based Sub-Channel Assignment}
In this section, the formulated sub-channel assignment problem in \eqr{saproblem} is solved with the given resource allocation solutions. Specifically, the optimal resource allocation for all devices assigned to all sub-channels can be obtained in Section IV, and therefore, this solution is treated as a preference list to construct a matching based sub-channel assignment algorithm. Note that some combinations of devices and sub-channels may not be feasible due to inability to satisfy the maximum time consumption constraint, and the proposed algorithm may tend to assign devices to the corresponding infeasible sub-channels to achieve lower energy consumption. In order to avoid this case, a large value is assigned as the energy consumption of these infeasible combinations, and any combination with this energy consumption will be removed from the final matching.
%%%%%%%%%%%%%%%%%%%%%%%%%%%%%%%%%%%%%%%%%%%%%%%%%
\subsection{Design of Matching based Algorithm}
At this stage, with the preference list setting, all devices in $\mathcal{S}_t$ can be assigned to sub-channels, and thus problem \eqr{saproblem} can be considered as a one-to-one matching $\Psi$ from $\mathcal{S}_t$ to $\mathcal{K}$, where $\mathcal{S}_t$ and $\mathcal{K}$ are two disjoint sets with the same size. In the resource allocation problem, it is indicated that the energy consumption of any device $n$ or sub-channel $k$ in matching $\Psi$ is independent of other players, and therefore, the utility of any player can be defined as follows:
\begin{equation}
U_i(\Psi) =E_{k,n}^{\mathrm{cp}}+E_{k,n}^{\mathrm{cm}}, \forall i \in \{n, k\}.
\end{equation}
Due to the fact that the device and sub-channel in a combination have the same utility, the intent of sub-channels can be omitted. Moreover, since each device is assigned to one sub-channel and each sub-channel is occupied by one device, if a device tends to establish a new matching, it needs to exchange with another device instead of joining the combination directly. That is, the considered matching is a swap matching, defined as follows:
\begin{definition}
From matching $\Psi$ with $\Psi(n)=k$ and $\Psi(n')=k'$, a swap matching $\Psi_n^{n'}$ represents an exchange of devices $n$ and $n'$, i.e.,
\begin{equation}
\Psi_n^{n'}=\Psi\backslash\{\{k, n\}, \{k',n'\}\} \cup \{\{k, n'\}, \{k',n\}\}.
\end{equation}
\end{definition}
As defined above, a swap matching means that two devices exchange their assigned sub-channels. Note that the motivation to form a swap matching is the reduction in energy consumption, which can be presented as follows:
\begin{equation}
\Psi\preceq_i\Psi_n^{n'} \Leftrightarrow U_i(\Psi) \le U_i(\Psi_n^{n'}), \forall i \in \{n, n'\},
\end{equation}
where $\Psi\preceq_i\Psi_n^{n'}$ indicates device $i$ prefers $\Psi_n^{n'}$ to $\Psi$. Moreover, symbol $\prec_i$ is also introduced to represent the strict preference of device $i$. The swap matching should be approved by all involved players, in which the utility of any player increases or remains unchanged. In this case, devices $n$ and $n'$ becomes a swap-blocking pair $(n, n')$, defined as follows:
\begin{definition}
$(n, n')$ is a swap-blocking pair if and only if $\Psi\prec_i\Psi_n^{n'}, \exists i\in \{n, n'\}$ and $\Psi\preceq_i\Psi_n^{n'}, \forall i\in \{n, n'\}$.
\end{definition}
Based on the definition of the swap-blocking pair, a matching based sub-channel assignment algorithm is presented in Algorithm \ref{malg}. In this algorithm, an initial matching is firstly obtained by randomly assigning all devices into all sub-channels. Afterwards, each device in turn operates on the remaining devices in order to find the swap-blocking pair. If any two devices can form a swap-blocking pair, their sub-channels are exchanged and the new matching is recorded. This algorithm is repeated until no new swap-blocking pair can be found in a complete cycle. Based on the finial matching provided by Algorithm \ref{malg}, the solution of sub-channel assignment problem \eqr{saproblem} can be obtained by removing all infeasible combinations.

\begin{algorithm}[t]
\caption{Matching based Algorithm}
\label{malg}
\begin{algorithmic}[1]
\STATE \textbf{Initialization:}
\STATE  Randomly match all players in $\mathcal{S}_t$ and $\mathcal{K}$ to obtain $\Psi$.
\STATE Set $\Psi_\alpha = 0$ and $\Psi_\beta = 1$.
\STATE \textbf{Main Loop:}
\IF{$\Psi_\alpha\neq \Psi_\beta$}
\STATE Set $\Psi_\alpha = \Psi$.
\FOR{$n\in\mathcal{S}_t$}
\STATE Device $n$ searches device $n'\in\mathcal{S}_t$, where $n\neq n'$.
\IF{$(n, n')$ is a swap-blocking pair}
\STATE Devices $n$ and $n'$ exchange sub-channels.
\STATE Matching $\Psi_{n'}^n$ is obtained.
\STATE Set $\Psi=\Psi_{n'}^n$.
\ENDIF
\ENDFOR
\STATE Set $\Psi_\beta = \Psi$.
\ENDIF
\end{algorithmic}
\end{algorithm}

%%%%%%%%%%%%%%%%%%%%%%%%%%%%%%%%%%%%%%%%%%%%%%%%%
\subsection{Properties Analysis}
In this subsection, the properties of the proposed matching based sub-channel assignment algorithm, including complexity, convergence, and stability, are analyzed.

\subsubsection{Complexity}
The computational complexity of the proposed algorithm is $\mathcal{O}(CK^2)$, where $C$ is the number of cycles. Specifically, during a complete cycle of the main loop, each device needs to test the viability of creating swap-blocking pairs with all other devices, and hence, for all $K$ devices, $K(K-1)$ times of calculations should be performed. With the given number of cycles $C$, the computational complexity can be expressed as $CK(K-1)$.

\subsubsection{Convergence}
From any initial matching, the proposed algorithm is guaranteed to converge to a final matching without swap-blocking pairs. It can be observed from Algorithm \ref{malg} that the matching is transformed due to the construction of swap-blocking pairs. Suppose $\Psi_a$ and $\Psi_b$ are two adjacent matching, where $a\neq b$, then there exists a swap-blocking pair in the transformation from $\Psi_a$ to $\Psi_b$. Based on the definition of a swap-blocking pair, it indicates that the utility of at least one device is strictly reduced while the utility of the other device is not increased. Moreover, for the devices that are not involved, their utility remains the same. As a result, the sum utility, or sum energy consumption, is strictly decreased with this transformation. With the given devices and sub-channels, there is a lower bound on the energy consumption, and therefore, the proposed algorithm can always converge to a final matching.

\subsubsection{Stability}
The proposed matching based sub-channel assignment algorithm is able to provide a two-side exchange stable solution, which is defined as follows:
\begin{definition}
A matching is two-side exchange stable if and only if no swap-blocking pair can be formed.
\end{definition}
According to the above definition, the final matching obtained by the proposed algorithm is alway two-side exchange stable, since the convergence analysis proves that there are no swap-blocking pairs in the final matching.
%%%%%%%%%%%%%%%%%%%%%%%%%%%%%%%%%%%%%%%%%%%%%%%%%
%%%%%%%%%%%%%%%%%%%%%%%%%%%%%%%%%%%%%%%%%%%%%%%%%
\section{Simulation Results}
The simulation results are presented to demonstrate the performance of age-weighted FedSGD and proposed solutions. In this simulation, the devices are randomly deployed in a disc with radius $R$, while the server is located in the center. The learning rate $\lambda$ is  $0.01$, bandwidth $B$ is $1$~MHz, noise power $\sigma^2$ is  $-174$~dBm, path loss exponent $a$ is $3.76$, power consumption coefficient $\kappa$ is $10^{-29}$, and cycles coefficient $\mu$ is $10^6$. To evaluate the learning performance, MNIST, CIFAR-10, and CIFAR-100 datasets are adopted with an SGD optimizer. For MNIST digit recognition tasks, a simple neural network is built with a $128$-neuron ReLu hidden layer and a softmax output layer. For CIFAR-10 image classification tasks, a neural network is constructed by stacking a $32$-filter $4\times 4$ 2D convolution (Conv2D) layers, a $2\times 2$ max pooling layer, a $128$-neuron ReLu hidden layer and a softmax output layer. For CIFAR-100 image classification tasks,, the neural network is constructed with a $128$-filter $4\times 4$ Conv2D layers, a $2\times 2$ max pooling layer, a $256$-neuron ReLu hidden layer and a softmax output layer. To compare the performance of the proposed KKT based resource allocation (denoted by KRA) and matching based sub-channel assignment (denoted by MSA), fixed resource allocation (denoted by FRA) and random sub-channel assignment (denoted by RSA) are respectively included as the baseline, where resource allocation coefficients are set as $\tau_{k,n}=\alpha_{k,n}=0.5, \forall k, n$ with FRA 1, and $\tau_{k,n}=\alpha_{k,n}=1, \forall k, n$ with FRA 2.

\begin{figure}[!t]
\centering{\includegraphics[width=100mm]{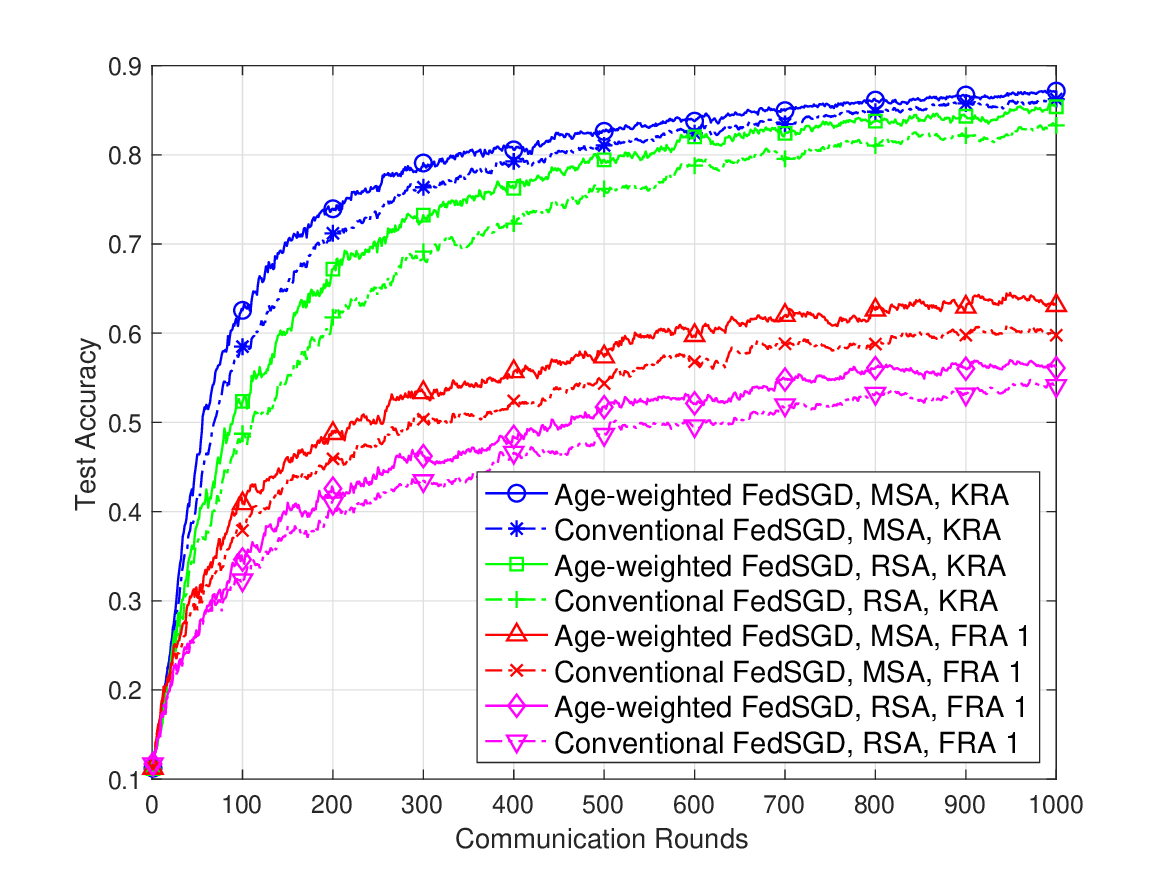}}
\caption{The convergence performance on the unbalanced MNIST dataset. $N=10$, $K=4$, $T_{n}^{\mathrm{max}}=5$~s, $P_n=10$~dBm, $C_n = 1$~GHz, $R=200$~m, and $D=10$~Mbits.}
\label{result1}
\vspace{-4mm}
\end{figure}

\begin{figure}[!t]
\centering{\includegraphics[width=100mm]{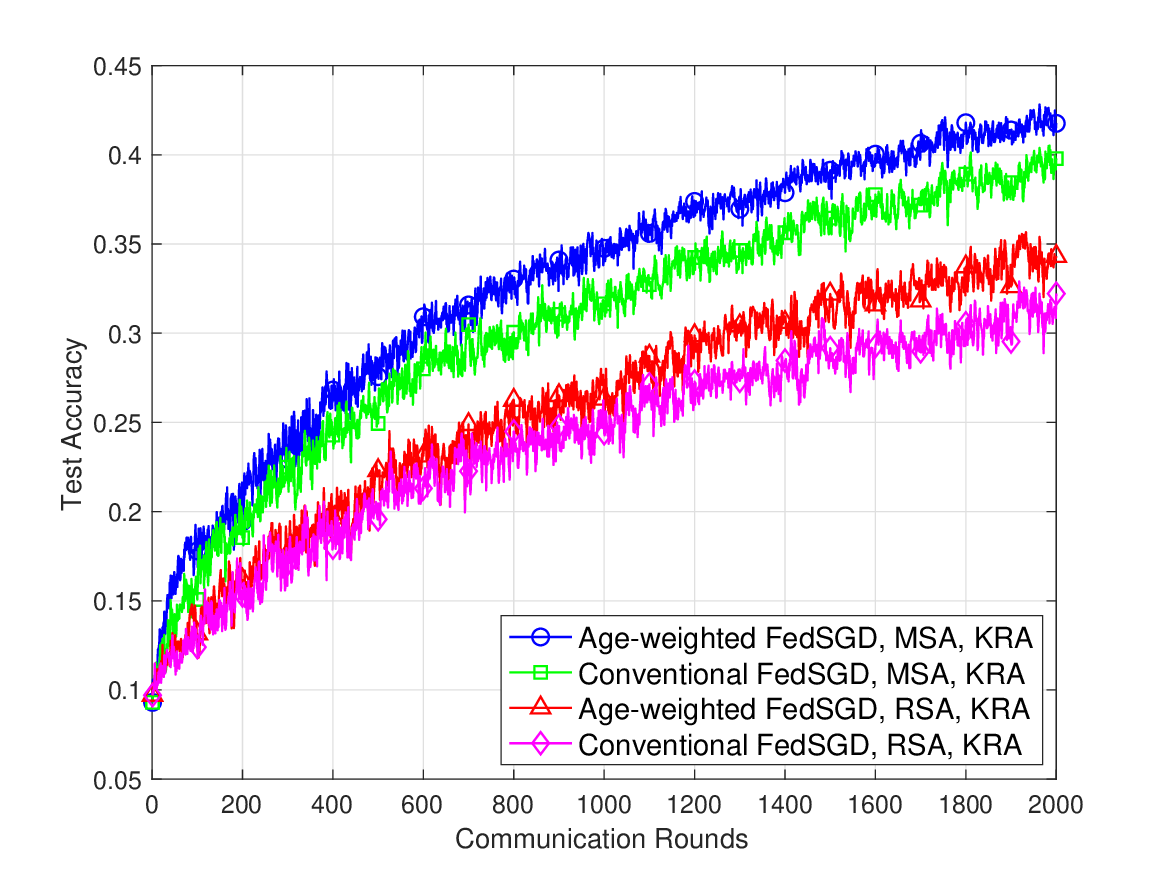}}
\caption{The convergence performance on the balanced CIFAR10 dataset. $N=10$, $K=5$, $T_{n}^{\mathrm{max}}=10$~s, $P_n=10$~dBm, $C_n = 1$~GHz, $R=200$~m, and $D=15$~Mbits.}
\label{result3}
\vspace{-4mm}
\end{figure}

In \fref{result1}, the MNIST digit recognition task with unbalanced non-IID data is adopted, where $9000$ training samples are randomly distributed to all devices, and the samples of each device belong to $1$ or $2$ classes. It demonstrates that with the same device selection results, the proposed age-weighted FedSGD is able to outperform conventional FedSGD with any sub-channel assignment and resource allocation schemes. Compared to RSA and FRA, MSA and KRA can also improve the performance of federated learning. Moreover, since the number of selected devices in each communication round is obviously reduced with FRA, there exist a large gap between FRA and KRA in achievable test accuracy.

The CIFAR-10 image classification task is employed in \fref{result3}, in which each device has $5000$ samples with the unique label. In this figure, the average number of selected devices is $2.4541$ and $3.7392$ for RSA and MSA, respectively. In other words, by utilizing the proposed sub-channel assignment strategy, the device availability is increased by $52\%$. As a result, for both age-weight FedSGD and conventional FedSGD, the learning performance is improved with the proposed solution. Moreover, it can be found from \fref{result3} that with the same device selection results, age-weighted FedSGD can achieve faster convergence rate compared to conventional FedSGD.

\begin{figure}[!t]
\centering{\includegraphics[width=100mm]{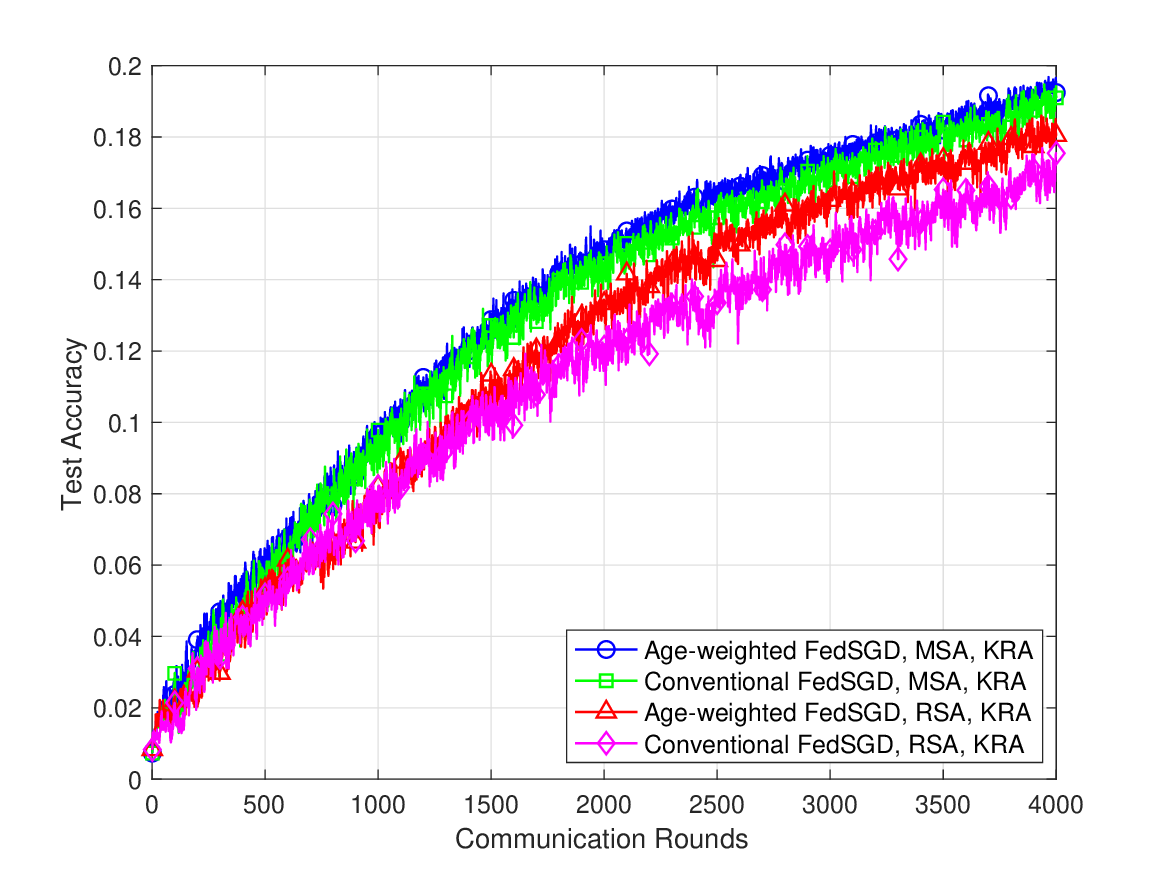}}
\caption{The convergence performance on the balanced CIFAR100 dataset. $N=50$, $K=20$, $T_{n}^{\mathrm{max}}=10$~s, $P_n=10$~dBm, $C_n = 1$~GHz, $R=200$~m, and $D=20$~Mbits.}
\label{result4}
\vspace{-4mm}
\end{figure}

\begin{figure}[!t]
\centering{\includegraphics[width=120mm]{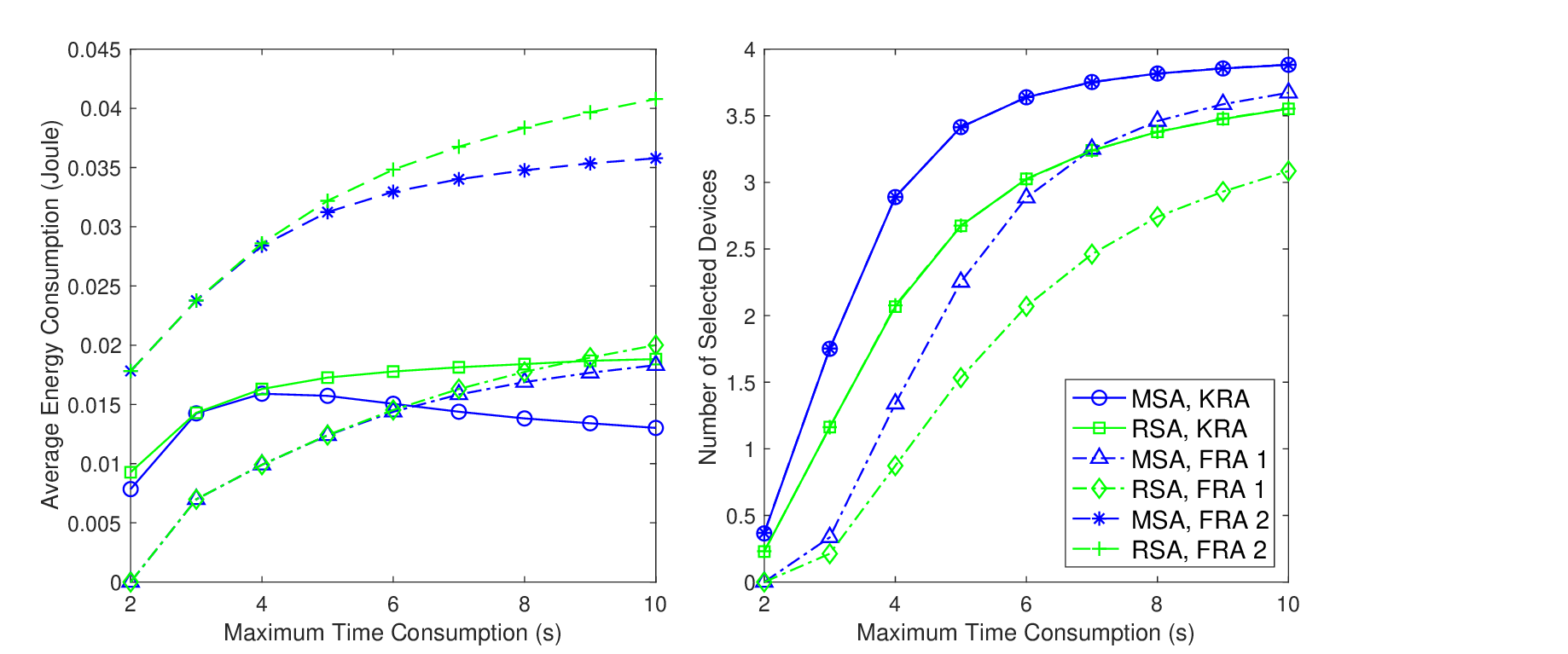}}
\caption{The impact of the maximum time consumption. $N=10$, $K=4$, $P_n=10$~dBm, $C_n = 1$~GHz, $R=200$~m, and $D=10$~Mbits.}
\label{resultt}
\vspace{-4mm}
\end{figure}

The balanced CIFAR-100 dataset is adopted in \fref{result4}, in which each device has $1000$ $2$-class samples. Due to the fact that the Non-IID degree in this figure is decreased compared to that in \fref{result1} and \fref{result3}, the improvement of age-weighted FedSGD with MSA is not significant, but it can still improve the learning performance, especially in the later stage of training. With RSA, the advantage of age-weighted FedSGD is obvious, and this is because the number of selected devices in each round is significantly reduced. Specifically, under this parameter setting, the device selection probabilities are $97.96\%$ and $65.45\%$ with MSA and RSA, respectively.

\begin{figure}[!t]
\centering{\includegraphics[width=120mm]{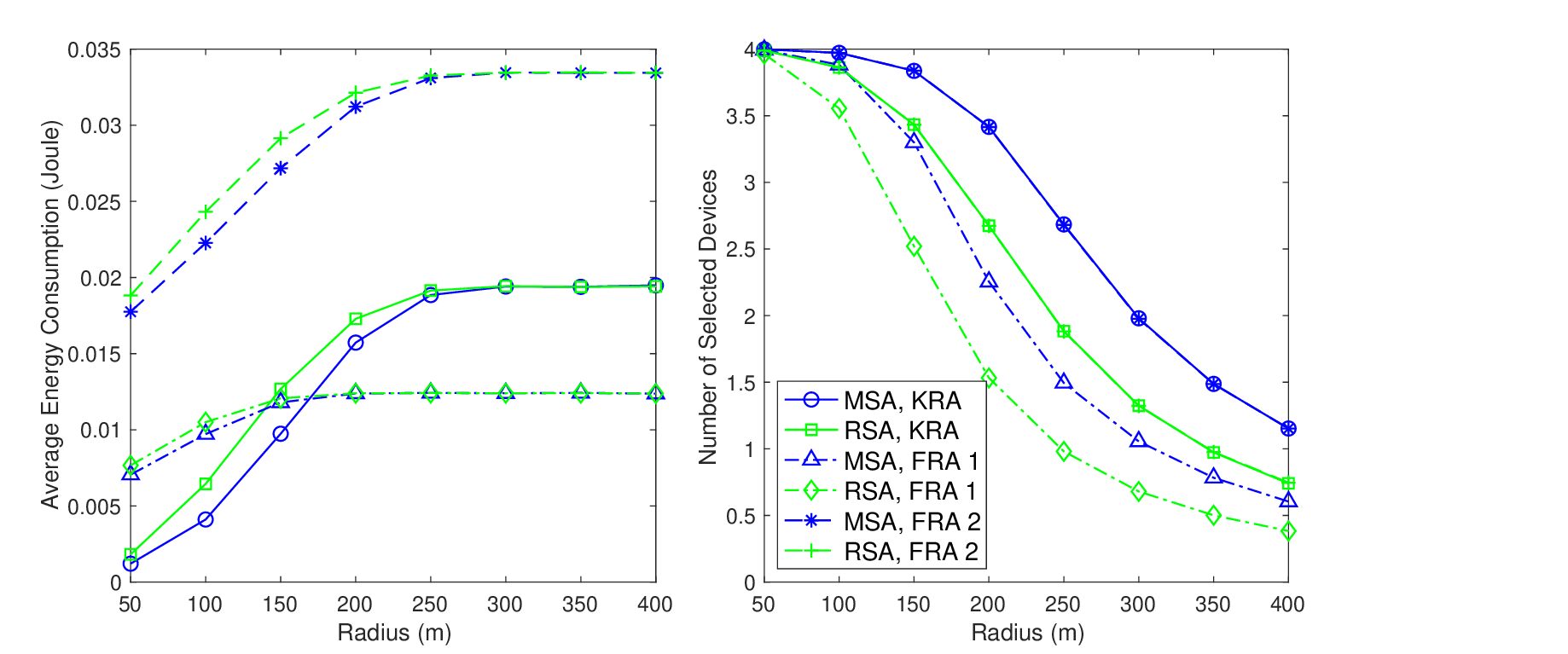}}
\caption{The impact of channel conditions. $N=10$, $K=4$, $T_{n}^{\mathrm{max}}=5$~s, $P_n=10$~dBm, $C_n = 1$~GHz, and $D=10$~Mbits.}
\label{resultr}
\vspace{-4mm}
\end{figure}

\begin{figure}[!t]
\centering{\includegraphics[width=120mm]{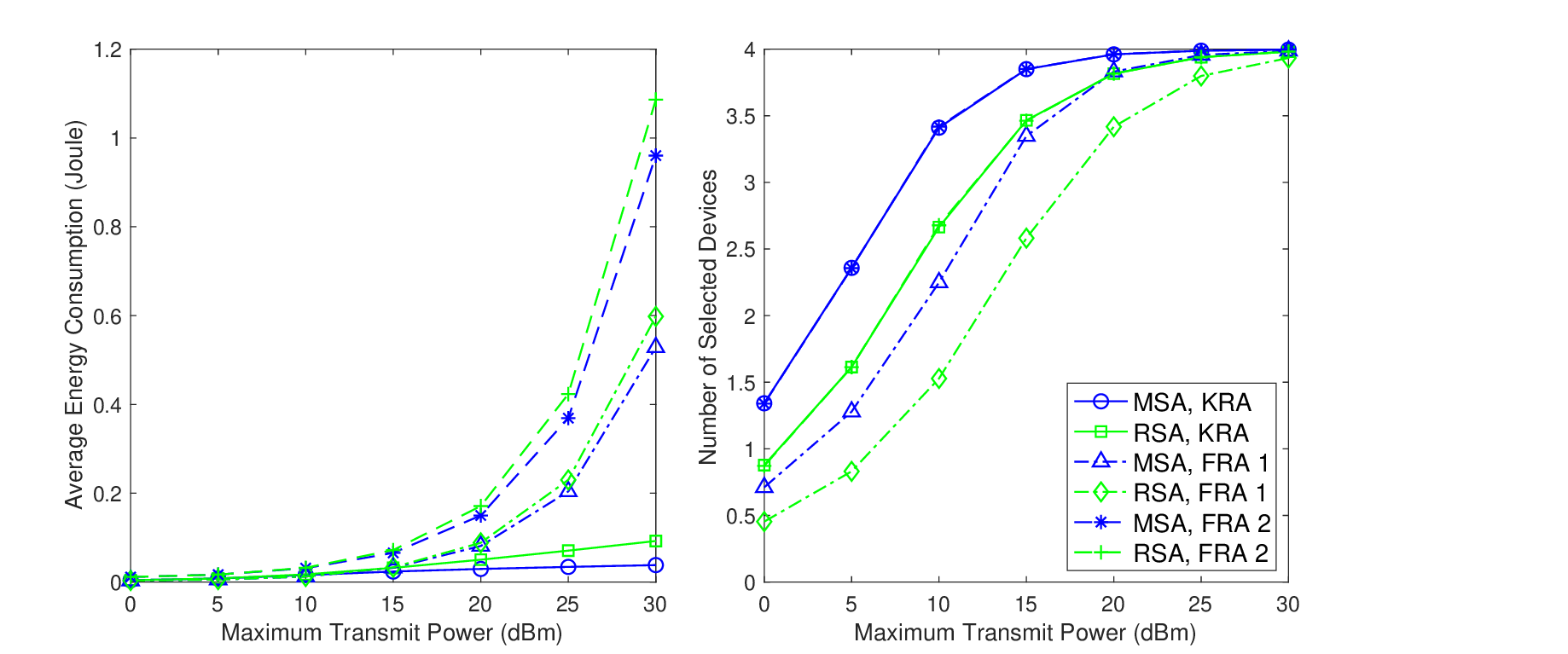}}
\caption{The impact of the maximum transmit power. $N=10$, $K=4$, $T_{n}^{\mathrm{max}}=5$~s, $C_n = 1$~GHz, $R=200$~m, and $D=10$~Mbits.}
\label{resultpn}
\vspace{-4mm}
\end{figure}

\begin{figure}[!t]
\centering{\includegraphics[width=120mm]{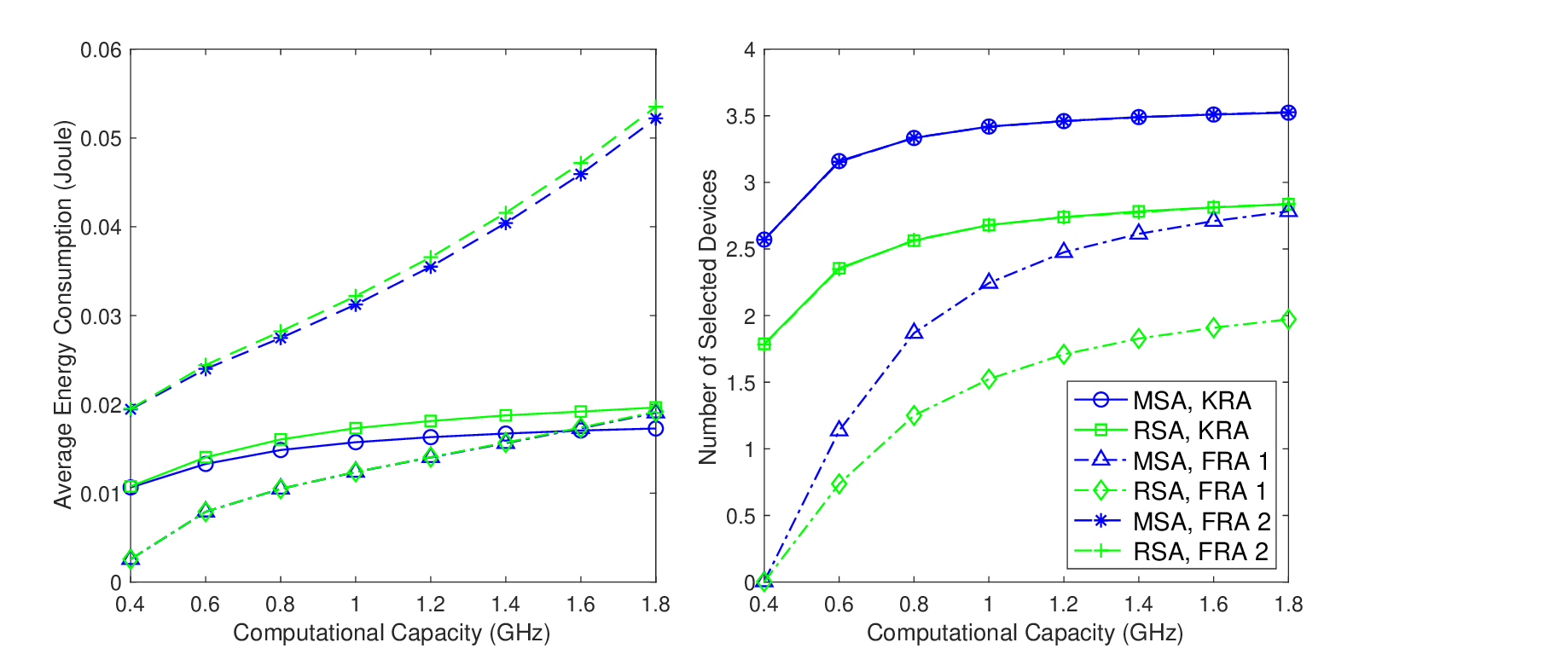}}
\caption{The impact of the computational capacity. $N=10$, $K=4$, $T_{n}^{\mathrm{max}}=5$~s, $P_n=10$~dBm, $R=200$~m, and $D=10$~Mbits.}
\label{resultcn}
\vspace{-4mm}
\end{figure}

The performance of the proposed solutions is shown in \fref{resultt} to \fref{resultpn}, where the maximum time consumption, radius, maximum transmit power, and computational capacity are respectively included to show its impact on the average energy consumption and the number of selected devices. It can be observed that compared to FRA 2, the obtained KKT based resource allocation solution can guarantee the same number of selected devices and achieve less energy consumption. Compared to FRA 1, KRA can reduce energy consumption and increase the number of selected devices, which explains the improved learning performance in \fref{result1} to \fref{result4}. Moreover, \fref{resultt} indicates that there is a trade-off between energy consumption and the number of selected devices. Specifically, when the maximum time consumption increases from $2$~s to $4$~s, the energy consumption is raised in order to significantly increase the number of selected devices, from $0.4$ to $3$. When the maximum time consumption is greater than $4$~s, the number of selected devices can be slowly increased, but the average energy consumption is reduced. In \fref{resultr}, the increase in radius can be regarded as the deterioration of channel conditions, and then it can be observed that there is an upper boundary in energy consumption. That is, when the radius is greater than $250$~m, the average energy consumption is maintained at a fixed level even though the number of selected devices continues to decrease. Similarly, an upper bound in the number of selected devices can be found in \fref{resultpn}. When the maximum transmit power is equal to $30$~dBm, the maximum device availability is achieve by all schemes, and the proposed solutions still consume minimum energy. Compared to the maximum transmit power, the impact of computational capacity is not significant, but the overall trend is the same, i.e., the energy consumption and the number of selected devices increase monotonically, as shown in \fref{resultcn}.

\begin{figure}[!t]
\centering{\includegraphics[width=100mm]{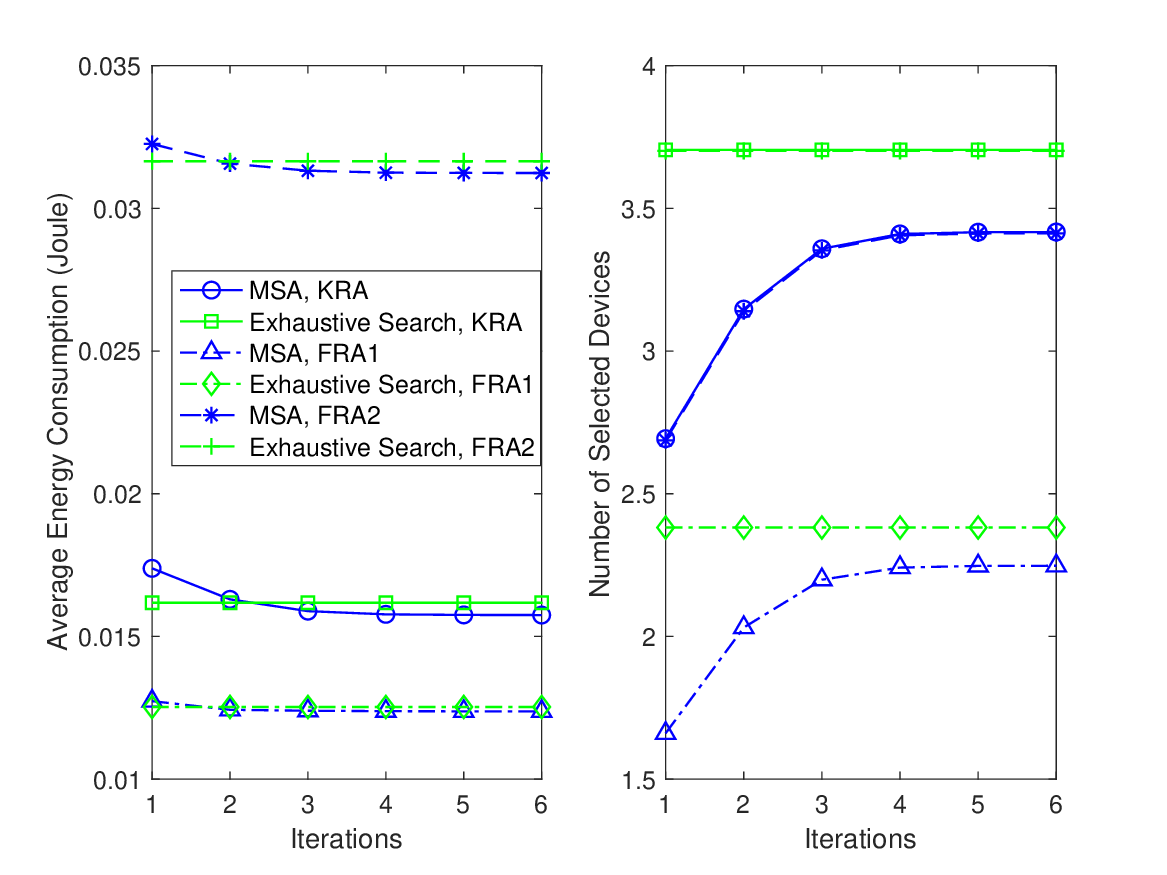}}
\caption{The convergence of the matching based sub-channel assignment algorithm. $N=10$, $K=4$, $T_{n}^{\mathrm{max}}=5$~s, $P_n=10$~dBm, $C_n = 1$~GHz, $R=200$~m, and $D=10$~Mbits.}
\label{resultmsa}
\vspace{-4mm}
\end{figure}

The convergence of the proposed sub-channel assignment algorithm is demonstrated in \fref{resultmsa}, where exhaustive search is included as the benchmark. In this figure, the exhaustive search is set up to find the combination that maximizes the number of selected devices while guaranteeing a low level of average energy consumption. Therefore, although the average energy consumption obtained by the exhaustive search is slightly larger than that of the proposed matching based algorithm, the number of selected devices is significantly increased. It can be observed that the proposed sub-channel assignment algorithm can achieve approximately $92\%$ performance of the global optimum within $4$ iterations. Compared to exhaustive search with complexity $\mathcal{O}(K!)$, it can be considered as a low complexity sub-optimal algorithm.  Furthermore, the figure verifies the properties of the proposed algorithm, including convergence and stability.
%%%%%%%%%%%%%%%%%%%%%%%%%%%%%%%%%%%%%%%%%%%%%%%%%
%%%%%%%%%%%%%%%%%%%%%%%%%%%%%%%%%%%%%%%%%%%%%%%%%
\section{Conclusions}
This paper investigates a wireless federated learning framework on non-IID datasets, where random device selection is exploited due to the limited number of sub-channels. By exploring the issue of conventional FedSGD in weight divergence, age-weighted FedSGD is designed to adjust the proportion of local gradients according to the previous state of devices. To further improve the learning performance, an energy consumption minimization problem is formulated, where the resource allocation solution and the sub-channel assignment algorithm are developed based on KKT conditions and matching theory, respectively. The superiority of designed age-weighted FedSGD and the effectiveness of the proposed resource allocation and sub-channel assignment strategies are demonstrated in the simulation results.
%%%%%%%%%%%%%%%%%%%%%%%%%%%%%%%%%%%%%%%%%%%%%%%%%
%%%%%%%%%%%%%%%%%%%%%%%%%%%%%%%%%%%%%%%%%%%%%%%%%
\section*{Appendix~A: Proof of Theorem~\ref{weightdivergence}}
According to \eqr{updatefed} and \eqr{updatetrue}, the weight divergence between random device selection and complete device selection can be expressed as follows:
\begin{equation}
\left\|\mathbf{w}^{\mathrm{(t+1)}}-\mathbf{w}_\mathrm{T}^{\mathrm{(t+1)}}\right\|= \left\|\mathbf{w}^{\mathrm{(t)}}\!-\!\mathbf{w}_\mathrm{T}^{\mathrm{(t)}}\!-\!\lambda\nabla F(\mathbf{w}^{\mathrm{(t)}}, \mathcal{S}_t)\!+\!\lambda\nabla F(\mathbf{w}_\mathrm{T}^{\mathrm{(t)}}, \mathcal{N})\right\|.
\end{equation}
From \eqr{error}, $\nabla F(\mathbf{w}^{\mathrm{(t)}}, \mathcal{S}_t)=\mathbf{e}^\mathrm{(t)}+\nabla F(\mathbf{w}^{\mathrm{(t)}}, \mathcal{N})$ can be obtained, and the above equation can be transformed as follows:
\begin{align}\nonumber
&\left\|\mathbf{w}^{\mathrm{(t+1)}}-\mathbf{w}_\mathrm{T}^{\mathrm{(t+1)}}\right\|\\\nonumber
= &\left\|\mathbf{w}^{\mathrm{(t)}}\!-\!\mathbf{w}_\mathrm{T}^{\mathrm{(t)}}\!-\!\lambda\mathbf{e}^\mathrm{(t)}\!-\!\lambda\!\left[\nabla F(\mathbf{w}^{\mathrm{(t)}}, \mathcal{N})\!-\!\nabla F(\mathbf{w}_\mathrm{T}^{\mathrm{(t)}}, \mathcal{N})\right]\right\|\\\nonumber
= &\left\|\mathbf{w}^{\mathrm{(1)}}\!\!-\!\mathbf{w}_\mathrm{T}^{\mathrm{(1)}}\!\!-\!\!\lambda\!\!\sum_{i=1}^t\!\mathbf{e}^{(i)}\!\!-\!\!\lambda\!\!\sum_{i=1}^t\!\left[\!\nabla F(\mathbf{w}^{(i)}, \mathcal{N})\!-\!\!\nabla F(\mathbf{w}_\mathrm{T}^{(i)}, \mathcal{N})\!\right]\!\right\|\\
\le &\left\|\mathbf{w}^{\mathrm{(1)}}\!-\!\mathbf{w}_\mathrm{T}^{\mathrm{(1)}}\right\|\!+\!\lambda\!\left\|\sum_{i=1}^t\mathbf{e}^{(i)}\right\|\!+\!\lambda\!\sum_{i=1}^t\left\|\nabla F(\mathbf{w}^{(i)}, \mathcal{N})\!-\!\nabla F(\mathbf{w}_\mathrm{T}^{(i)}, \mathcal{N})\right\|.
\end{align}
Based on Assumption \ref{assume1}, the following inequality holds:
\begin{equation}\label{astep1}
\left\|\mathbf{w}^{\mathrm{(t+1)}}-\mathbf{w}_\mathrm{T}^{\mathrm{(t+1)}}\right\|\le \left\|\mathbf{w}^{\mathrm{(1)}}\!-\!\mathbf{w}_\mathrm{T}^{\mathrm{(1)}}\right\|\!+\!\lambda\left\|\sum_{i=1}^t\mathbf{e}^{(i)}\right\|\!+\!\lambda L\!\sum_{i=1}^t\left\|\mathbf{w}^{(i)}\!-\!\mathbf{w}_\mathrm{T}^{(i)}\right\|.
\end{equation}
With the similar way, the following inequality can be obtained:
\begin{equation}
\lambda L\!\left\|\mathbf{w}^{\mathrm{(t)}}-\mathbf{w}_\mathrm{T}^{\mathrm{(t)}}\right\|\le \lambda L\!\left\|\mathbf{w}^{\mathrm{(1)}}\!-\!\mathbf{w}_\mathrm{T}^{\mathrm{(1)}}\right\|\!+\!\lambda^2 L\!\left\|\sum_{i=1}^{t-1}\mathbf{e}^{(i)}\right\|\!+\!(\lambda L)^2\!\sum_{i=1}^{t-1}\left\|\mathbf{w}^{(i)}\!-\!\mathbf{w}_\mathrm{T}^{(i)}\right\|.
\end{equation}
By substituting the above inequality, \eqr{astep1} can be rewritten as follows:
\begin{align}\nonumber
&\left\|\mathbf{w}^{\mathrm{(t+1)}}-\mathbf{w}_\mathrm{T}^{\mathrm{(t+1)}}\right\|\\
\le &(1\!+\!\lambda L)\!\left\|\mathbf{w}^{\mathrm{(1)}}\!-\!\mathbf{w}_\mathrm{T}^{\mathrm{(1)}}\right\|\!+\!\lambda\left\|\sum_{i=1}^t\mathbf{e}^{(i)}\right\|\!+\!\lambda^2 L\left\|\sum_{i=1}^{t-1}\mathbf{e}^{(i)}\right\|\!+\!\lambda L(1\!+\!\lambda L)\sum_{i=1}^{t-1}\left\|\mathbf{w}^{(i)}\!-\!\mathbf{w}_\mathrm{T}^{(i)}\right\|.
\label{astep2}
\end{align}
Similarly, $\lambda L(1\!+\!\lambda L)\!\left\|\mathbf{w}^{\mathrm{(t-1)}}-\mathbf{w}_\mathrm{T}^{\mathrm{(t-1)}}\right\|$ in the above inequality can be transformed to
\begin{align}\nonumber
&\lambda L(1\!+\!\lambda L)\!\left\|\mathbf{w}^{\mathrm{(t-1)}}-\mathbf{w}_\mathrm{T}^{\mathrm{(t-1)}}\right\|\\
\le &\lambda L(1\!+\!\lambda L)\!\left\|\mathbf{w}^{\mathrm{(1)}}\!-\!\mathbf{w}_\mathrm{T}^{\mathrm{(1)}}\right\|\!+\!\lambda^2 L(1\!+\!\lambda L)\!\left\|\sum_{i=1}^{t-2}\mathbf{e}^{(i)}\right\|\!+\!(\lambda L)^2(1\!+\!\lambda L)\!\sum_{i=1}^{t-2}\left\|\mathbf{w}^{(i)}\!-\!\mathbf{w}_\mathrm{T}^{(i)}\right\|,
\end{align}
and included in \eqr{astep2} to obtain the following inequality:
\begin{align}\nonumber
&\left\|\mathbf{w}^{\mathrm{(t+1)}}-\mathbf{w}_\mathrm{T}^{\mathrm{(t+1)}}\right\|\\
\le &(1\!+\!\lambda L)^2\!\left\|\mathbf{w}^{\mathrm{(1)}}\!-\!\mathbf{w}_\mathrm{T}^{\mathrm{(1)}}\right\|\!\!+\!\!\lambda\!\left\|\sum_{i=1}^t\mathbf{e}^{(i)}\right\|\!\!+\!\!\lambda^2 L\!\left\|\sum_{i=1}^{t-1}\mathbf{e}^{(i)}\right\|\\\nonumber
&+\!\lambda^2 L(1\!+\!\lambda L)\!\left\|\sum_{i=1}^{t-2}\mathbf{e}^{(i)}\right\|\!+\!\lambda L(1\!+\!\lambda L)^2\!\sum_{i=1}^{t-2}\!\left\|\mathbf{w}^{(i)}\!-\!\mathbf{w}_\mathrm{T}^{(i)}\right\|.
\end{align}
By induction, the weight divergence between random device selection and complete device selection can be presented as follows:
\begin{align}\nonumber
&\left\|\mathbf{w}^{\mathrm{(t+1)}}\!-\!\mathbf{w}_\mathrm{T}^{\mathrm{(t+1)}}\right\|\\
\le &(1\!+\!\lambda L)^t\!\left\|\mathbf{w}^{\mathrm{(1)}}\!-\!\mathbf{w}_\mathrm{T}^{\mathrm{(1)}}\right\|\!\!+\!\!\lambda\!\left\|\sum_{i=1}^{t}\mathbf{e}^{(i)}\right\|\!\!+\!\!\lambda^2 L\!\left\|\sum_{i=1}^{t-1}\mathbf{e}^{(i)}\right\|\\\nonumber
&+\!\!\lambda^2 L(1\!+\!\lambda L)\!\left\|\sum_{i=1}^{t-2}\!\mathbf{e}^{(i)}\!\right\|\!\!+\!\!\lambda^2 L(1\!+\!\lambda L)^2\!\left\|\sum_{i=1}^{t-3}\!\mathbf{e}^{(i)}\!\right\|\!+\!\cdots.
\end{align}
The above inequality can be rewritten as follows:
\begin{equation}
\left\|\mathbf{w}^{\mathrm{(t+1)}}-\mathbf{w}_\mathrm{T}^{\mathrm{(t+1)}}\right\|\le (1\!+\!\lambda L)^t\left\|\mathbf{w}^{\mathrm{(1)}}\!-\!\mathbf{w}_\mathrm{T}^{\mathrm{(1)}}\right\|\!+\!\lambda\left\|\sum_{i=1}^{t}\mathbf{e}^{(i)}\right\|\!+\!\lambda^2L\!\sum_{j=1}^{t-1}(1\!+\!\lambda L)^{j-1}\!\left\|\sum_{i=1}^{t-j}\!\mathbf{e}^{(i)}\right\|,
\end{equation}
and the proof is completed.\QEDA
%%%%%%%%%%%%%%%%%%%%%%%%%%%%%%%%%%%%%%%%%%%%%%%%%
%%%%%%%%%%%%%%%%%%%%%%%%%%%%%%%%%%%%%%%%%%%%%%%%%
\section*{Appendix~B: Proof of Theorem~\ref{convergence}}
Based on \eqr{cond1} and \eqr{cond2}, the upper bound of the convergence rate can be proved through a similar approach to that in \cite{kaidi2023fl2}. In order to derive the expression of $\mathbb{E}\!\left[\|\mathbf{g}^{(i)}\|^2\right]$, the gradient of the global loss, i.e., $\nabla G(\mathbf{w}^\mathrm{(t)},\mathcal{S}_t)$, can be viewed as ratio estimation, as follows:
\begin{equation}
\nabla G(\mathbf{w}^\mathrm{(t)},\mathcal{S}_t)=\frac{\frac{1}{|\mathcal{S}_t|}\sum_{n\in\mathcal{N}}x_n^\mathrm{(t)}\omega_n^\mathrm{(t)}\beta_n\nabla f_n(\mathbf{w}^\mathrm{(t)})}{\frac{1}{|\mathcal{S}_t|}\sum_{n\in\mathcal{N}}x_n^\mathrm{(t)}\beta_n}\triangleq \frac{\bar{y}_{\mathcal{S}_t}}{\bar{x}_{\mathcal{S}_t}},
\end{equation}
where $x_n^\mathrm{(t)}$ is a binary variable to indicate the device selection result of device $n$ in round $t$. In particular, $x_n^\mathrm{(t)}=1$ indicates that device $n$ is selected in round $t$, i.e., $n\in \mathcal{S}_t$;  $x_n^\mathrm{(t)}=0$ otherwise. Since random device selection is adopted, in any communication round $t$, the probability of selecting device $n$ from $\mathcal{N}$ is given by
\begin{equation}
\mathbb{E}[x_n^\mathrm{(t)}] = P(x_n^\mathrm{(t)}=1)=\frac{|\mathcal{S}_t|}{N}.
\end{equation}
Similarly, the gradient of the global loss with complete device selection, i.e., $\nabla F(\mathbf{w}^\mathrm{(t)},\mathcal{N})$, can be expressed as follows:
\begin{equation}
\nabla F(\mathbf{w}^\mathrm{(t)},\mathcal{N})=\frac{\frac{1}{N}\sum_{n\in\mathcal{N}}\beta_n\nabla f_n(\mathbf{w}^\mathrm{(t)})}{\frac{1}{N}\sum_{n\in\mathcal{N}}\beta_n}\triangleq \frac{\bar{y}_{\mathcal{N}}}{\bar{x}_{\mathcal{N}}}.
\end{equation}
At this stage, the following equations can be obtained:
\begin{equation}
\mathbb{E}[\bar{x}_{\mathcal{S}_t}] = \bar{x}_\mathcal{N},
\end{equation}
and
\begin{equation}
\mathbb{E}[\bar{y}_{\mathcal{S}_t}] = \bar{y}_\mathcal{N}.
\end{equation}
Then, $\mathbb{E}\left[\|\mathbf{g}^{\mathrm{(t)}}\|^2\right]$ can be rewritten as follows:
\begin{equation}
\mathbb{E}\left[\|\mathbf{g}^{\mathrm{(t)}}\|^2\right]=\frac{1}{(\frac{1}{N}\!\sum_{n\in\mathcal{N}}\!\beta_n)^2}\mathbb{E}\!\left[\left\|\bar{y}_{\mathcal{S}_t}-\bar{x}_{\mathcal{S}_t}\frac{\bar{y}_{\mathcal{N}}}{\bar{x}_{\mathcal{N}}}\right\|^2\right].
\end{equation}
The rest of the proof can be obtained by referring to \cite{kaidi2023fl2}.\QEDA
%%%%%%%%%%%%%%%%%%%%%%%%%%%%%%%%%%%%%%%%%%%%%%%%%
%%%%%%%%%%%%%%%%%%%%%%%%%%%%%%%%%%%%%%%%%%%%%%%%%
\section*{Appendix~C: Proof of Proposition~\ref{closedform}}
By calculating the partial derivatives of \eqr{lagrangian} and setting them to zero, the following equations can be obtained:
\begin{subequations}
\begin{empheq}[left=\empheqlbrace]{align}
&-\frac{2\kappa\mu\beta_n C_n^2}{(x_1^*)^ 3}+\lambda_1\frac{\mu\beta_n}{C_n}-\lambda_2=0,\\
&-\frac{\ln(2)D 2^\frac{1}{Bx_2^*}}{|h_{k,n}|^2Bx_2^*}+\frac{D(2^\frac{1}{Bx_2^*}-1)}{|h_{k,n}|^2}+\lambda_1 D-\lambda_3=0.
\end{empheq}
\label{kkt}
\end{subequations}\vspace{-2mm}\\
Moreover, the following conditions should be satisfied:
\begin{subequations}
\begin{empheq}[left=\empheqlbrace]{align}
\mu \beta_n C_n^{-1}x_1^*+Dx_2^*-T_{n}^{\mathrm{max}}\le0,\\
1-x_1^*\le0,\\
\upsilon_1-x_2^*\le 0,\\
\lambda_1(\mu \beta_n C_n^{-1}x_1^*+Dx_2^*-T_{n}^{\mathrm{max}})=0,\\
\lambda_2(1-x_1^*)=0,\\
\lambda_3(\upsilon_1-x_2^*)=0,\\
\lambda_i\!\ge\!0, \forall i \in\{1,2,3\},
\end{empheq}
\label{kktconditions}
\end{subequations}\vspace{-2mm}\\
where $\upsilon_1$ is defined in \eqr{vdefine}. If $\lambda_1 = 0$, the following equation can be obtained from (\ref{kkt}a):
\begin{equation}
-2\kappa\mu\beta_n C_n^2(x_1^*)^{-3}=\lambda_2.
\end{equation}
Since $\lambda_2\ge 0$, the above function conflicts with (\ref{kktconditions}b), and hence, $\lambda_1>0$ always holds. Therefore, it can be obtained from (\ref{kktconditions}d) that the following equation is always satisfied:
\begin{equation}\label{optimum}
\mu \beta_n C_n^{-1}x_1^*+Dx_2^*-T_{n}^{\mathrm{max}}=0.
\end{equation}
At this stage, based on the different values of $\lambda_2$ and $\lambda_3$, four cases need to be discussed.

1) If $\lambda_2>0$ and $\lambda_3>0$, $x_1^*=1$ and $x_2^*=\upsilon_1$ can be obtained from (\ref{kktconditions}e) and (\ref{kktconditions}f), respectively. In this case, the following condition can be derived from \eqr{optimum}:
\begin{equation}
\mu \beta_n C_n^{-1}+D\upsilon_1=T_{n}^{\mathrm{max}}.
\end{equation}

2) If $\lambda_2>0$ and $\lambda_3=0$, $x_1^*=1$, and it can be obtained from \eqr{optimum} that $x_2^*=(T_{n}^{\mathrm{max}}-\mu\beta_nC_n^{-1})/D$.  In this case, (\ref{kktconditions}c) should be considered, and the condition becomes
\begin{equation}\label{fcondition1}
\mu \beta_n C_n^{-1}+D\upsilon_1<T_{n}^{\mathrm{max}}.
\end{equation}
Note that the case $\lambda_3>0$ has been considered, and thus the equality condition is removed. Since $\lambda_3=0$, the following inequality can be obtained from (\ref{kkt}b):
\begin{equation}\label{lambda11}
\lambda_1=\frac{\ln(2)2^\frac{1}{Bx_2^*}}{|h_{k,n}|^2Bx_2^*}-\frac{2^\frac{1}{Bx_2^*}-1}{|h_{k,n}|^2}>0.
\end{equation}
Substituting the above equation into (\ref{kkt}a), it becomes
\begin{equation}
\left(\frac{\ln(2)2^\frac{1}{Bx_2^*}}{|h_{k,n}|^2Bx_2^*}\!-\!\frac{2^\frac{1}{Bx_2^*}\!-\!1}{|h_{k,n}|^2}\right)\frac{\mu\beta_n}{C_n}\!-\!2\kappa\mu\beta_n C_n^2=\lambda_2>0,
\end{equation}
and the following inequality can be obtained:
\begin{equation}
\frac{\ln(2)2^\frac{1}{Bx_2^*}}{|h_{k,n}|^2Bx_2^*}\!-\!\frac{2^\frac{1}{Bx_2^*}\!-\!1}{|h_{k,n}|^2}-2\kappa C_n^3>0.
\end{equation}
It is indicated that the above inequality includes \eqr{lambda11}, and thus condition \eqr{lambda11} can be omitted. By including the expression of $x_2^*$, the condition in this case is given by
\begin{equation}
D\upsilon_2\ln(2)2^{D\upsilon_2}\!-\!2^{D\upsilon_2}\!+\!1\!-\!2\kappa C_n^3|h_{k,n}|^2>0.
\end{equation}

3) If $\lambda_2=0$ and $\lambda_3>0$, by including $x_2^*=\upsilon_1$ to \eqr{optimum}, the expression of $x_1^*$ can be presented as follows:
\begin{equation}
x_1^*=(T_{n}^{\mathrm{max}}-D\upsilon_1)C_n(\mu\beta_n)^{-1},
\end{equation}
and (\ref{kktconditions}b) can be rewritten as \eqr{fcondition1}, where the equality condition is removed as it holds for the case $\lambda_2>0$. From (\ref{kkt}a), the following condition can be obtained:
\begin{equation}\label{lambda12}
\lambda_1=2\kappa C_n^3(x_1^*)^{-3}>0,
\end{equation}
which can be rewritten as follows:
\begin{equation}
D\upsilon_1<T_{n}^{\mathrm{max}}.
\end{equation}
The above inequality is always satisfied with inequality \eqr{fcondition1}. Moreover, the equation in \eqr{lambda12} can be substituted into (\ref{kkt}b), as shown in follows:
\begin{equation}
-\frac{D\ln(2) 2^\frac{1}{Bx_2^*}}{|h_{k,n}|^2Bx_2^*}\!+\!\frac{D(2^\frac{1}{Bx_2^*}\!-\!1)}{|h_{k,n}|^2}\!+\!\frac{2\kappa C_n^3 D}{(x_1^*)^3}=\lambda_3>0.
\end{equation}
By including the obtained solutions of $x_1^*$ and $x_2^*$, it becomes
\begin{equation}
2^\frac{1}{B\upsilon_1}\!-\!1\!-\!\frac{1}{B\upsilon_1}\!\ln(2)2^\frac{1}{B\upsilon_1}\!+\!\frac{2\kappa(\mu\beta_n)^3|h_{k,n}|^2}{(T_{n}^{\mathrm{max}}\!-\!D\upsilon_1)^3}>0.
\end{equation}

4) If $\lambda_2=0$ and $\lambda_3=0$, the following equation can be obtained from (\ref{kkt}a) and (\ref{kkt}b):
\begin{equation}
\frac{2\kappa C_n^3}{(x_1^*)^3}= \frac{\ln(2)2^\frac{1}{Bx_2^*}}{B|h_{k,n}|^2x_2^*}-\frac{2^\frac{1}{Bx_2^*}-1}{|h_{k,n}|^2}.
\end{equation}
Moreover, by including \eqr{optimum}, the solution can be obtained by solving the following equations:
\begin{equation}
\left\{\begin{array}{ll}
&\dfrac{2\kappa C_n^3}{(x_1^*)^3}= \dfrac{\ln(2)2^\frac{1}{Bx_2^*}}{B|h_{k,n}|^2x_2^*}-\dfrac{2^\frac{1}{Bx_2^*}-1}{|h_{k,n}|^2},\vspace{1mm}\\
&\mu \beta_n C_n^{-1}x_1^*+Dx_2^*-T_{n}^{\mathrm{max}}=0.
\end{array}\right.
\end{equation}
This proposition is proved.\QEDA
%%%%%%%%%%%%%%%%%%%%%%%%%%%%%%%%%%%%%%%%%%%%%%%%%
%%%%%%%%%%%%%%%%%%%%%%%%%%%%%%%%%%%%%%%%%%%%%%%%%
\bibliographystyle{IEEEtran}
\bibliography{KaidisBib}

% Generated by IEEEtran.bst, version: 1.14 (2015/08/26)
\begin{thebibliography}{10}
\providecommand{\url}[1]{#1}
\csname url@samestyle\endcsname
\providecommand{\newblock}{\relax}
\providecommand{\bibinfo}[2]{#2}
\providecommand{\BIBentrySTDinterwordspacing}{\spaceskip=0pt\relax}
\providecommand{\BIBentryALTinterwordstretchfactor}{4}
\providecommand{\BIBentryALTinterwordspacing}{\spaceskip=\fontdimen2\font plus
\BIBentryALTinterwordstretchfactor\fontdimen3\font minus \fontdimen4\font\relax}
\providecommand{\BIBforeignlanguage}[2]{{%
\expandafter\ifx\csname l@#1\endcsname\relax
\typeout{** WARNING: IEEEtran.bst: No hyphenation pattern has been}%
\typeout{** loaded for the language `#1'. Using the pattern for}%
\typeout{** the default language instead.}%
\else
\language=\csname l@#1\endcsname
\fi
#2}}
\providecommand{\BIBdecl}{\relax}
\BIBdecl

\bibitem{chen2020flm}
M.~Chen, H.~V. Poor, W.~Saad, and S.~Cui, ``Wireless communications for collaborative federated learning,'' \emph{{IEEE} Commun. Mag.}, vol.~58, no.~12, pp. 48--54, 2020.

\bibitem{mcmahan2017fl}
B.~McMahan, E.~Moore, D.~Ramage, S.~Hampson, and B.~A. y~Arcas, ``Communication-efficient learning of deep networks from decentralized data,'' in \emph{Artificial intelligence and statistics}.\hskip 1em plus 0.5em minus 0.4em\relax PMLR, 2017, pp. 1273--1282.

\bibitem{konevcny2016federated2}
J.~Kone{\v{c}}n{\`y}, H.~B. McMahan, F.~X. Yu, P.~Richt{\'a}rik, A.~T. Suresh, and D.~Bacon, ``Federated learning: Strategies for improving communication efficiency,'' \emph{arXiv preprint arXiv:1610.05492}, 2016.

\bibitem{savazzi2021fl}
S.~Savazzi, M.~Nicoli, M.~Bennis, S.~Kianoush, and L.~Barbieri, ``Opportunities of federated learning in connected, cooperative, and automated industrial systems,'' \emph{{IEEE} Commun. Mag.}, vol.~59, no.~2, pp. 16--21, 2021.

\bibitem{chen2021flm2}
M.~Chen, D.~G{\"{u}}nd{\"{u}}z, K.~Huang, W.~Saad, M.~Bennis, A.~V. Feljan, and H.~V. Poor, ``Distributed learning in wireless networks: Recent progress and future challenges,'' \emph{{IEEE} J. Sel. Areas Commun.}, vol.~39, no.~12, pp. 3579--3605, 2021.

\bibitem{yang2022fl}
Z.~Yang, M.~Chen, K.-K. Wong, H.~V. Poor, and S.~Cui, ``Federated learning for {6G}: Applications, challenges, and opportunities,'' \emph{Engineering}, vol.~8, pp. 33--41, 2022.

\bibitem{nishio2019fl}
T.~Nishio and R.~Yonetani, ``Client selection for federated learning with heterogeneous resources in mobile edge,'' in \emph{ICC 2019 - 2019 IEEE International Commun. Conf. (ICC)}, 2019, pp. 1--7.

\bibitem{fu2023fl}
L.~Fu, H.~Zhang, G.~Gao, M.~Zhang, and X.~Liu, ``Client selection in federated learning: Principles, challenges, and opportunities,'' \emph{IEEE Internet of Things Journal}, vol.~10, no.~24, pp. 21\,811--21\,819, 2023.

\bibitem{yang2020ds}
K.~Yang, T.~Jiang, Y.~Shi, and Z.~Ding, ``Federated learning via over-the-air computation,'' \emph{{IEEE} Trans. Wireless Commun.}, vol.~19, no.~3, pp. 2022--2035, 2020.

\bibitem{chen2021fl1}
M.~Chen, Z.~Yang, W.~Saad, C.~Yin, H.~V. Poor, and S.~Cui, ``A joint learning and communications framework for federated learning over wireless networks,'' \emph{{IEEE} Trans. Wireless Commun.}, vol.~20, no.~1, 2021.

\bibitem{guo2022fl}
W.~Guo, R.~Li, C.~Huang, X.~Qin, K.~Shen, and W.~Zhang, ``Joint device selection and power control for wireless federated learning,'' \emph{{IEEE} J. Sel. Areas Commun.}, vol.~40, no.~8, pp. 2395--2410, 2022.

\bibitem{ribero2023fl}
M.~Ribero, H.~Vikalo, and G.~de~Veciana, ``Federated learning under intermittent client availability and time-varying communication constraints,'' \emph{IEEE J. Sel. Top. Signal Process.}, vol.~17, no.~1, pp. 98--111, 2023.

\bibitem{konevcny2016federated}
J.~Kone{\v{c}}n{\`y}, H.~B. McMahan, D.~Ramage, and P.~Richt{\'a}rik, ``Federated optimization: Distributed machine learning for on-device intelligence,'' \emph{arXiv preprint arXiv:1610.02527}, 2016.

\bibitem{gafni2022flm}
T.~Gafni, N.~Shlezinger, K.~Cohen, Y.~C. Eldar, and H.~V. Poor, ``Federated learning: A signal processing perspective,'' \emph{{IEEE} Signal Process. Mag.}, vol.~39, no.~3, pp. 14--41, 2022.

\bibitem{zhao2018federated}
Y.~Zhao, M.~Li, L.~Lai, N.~Suda, D.~Civin, and V.~Chandra, ``Federated learning with non-iid data,'' \emph{arXiv preprint arXiv:1806.00582}, 2018.

\bibitem{mohanmmad2021fl}
M.~M. Amiri, D.~Gündüz, S.~R. Kulkarni, and H.~V. Poor, ``Convergence of update aware device scheduling for federated learning at the wireless edge,'' \emph{{IEEE} Trans. Wireless Commun.}, vol.~20, no.~6, pp. 3643--3658, 2021.

\bibitem{luo2022fl}
B.~Luo, W.~Xiao, S.~Wang, J.~Huang, and L.~Tassiulas, ``Tackling system and statistical heterogeneity for federated learning with adaptive client sampling,'' in \emph{IEEE INFOCOM 2022 - IEEE Conference on Computer Communications}, 2022, pp. 1739--1748.

\bibitem{cho2022towards}
Y.~J. Cho, J.~Wang, and G.~Joshi, ``Towards understanding biased client selection in federated learning,'' in \emph{International Conference on Artificial Intelligence and Statistics}.\hskip 1em plus 0.5em minus 0.4em\relax PMLR, 2022, pp. 10\,351--10\,375.

\bibitem{kaidi2023fl2}
K.~Wang, Z.~Ding, D.~K.~C. So, and Z.~Ding, ``Age-of-information minimization in federated learning based networks with {Non-IID} dataset,'' \emph{{IEEE} Trans. Wireless Commun.}, pp. 1--1, 2024.

\bibitem{li2020fl}
T.~Li, A.~K. Sahu, A.~Talwalkar, and V.~Smith, ``Federated learning: Challenges, methods, and future directions,'' \emph{{IEEE} Signal Process. Mag.}, vol.~37, no.~3, pp. 50--60, 2020.

\bibitem{yang2021fl}
Z.~Yang, M.~Chen, W.~Saad, C.~S. Hong, and M.~Shikh-Bahaei, ``Energy efficient federated learning over wireless communication networks,'' \emph{{IEEE} Trans. Wireless Commun.}, vol.~20, no.~3, pp. 1935--1949, 2021.

\bibitem{wang2023fl}
T.~Wang, N.~Huang, Y.~Wu, and T.~Q.~S. Quek, ``Energy-efficient wireless federated learning: A secrecy oriented design via sequential artificial jamming,'' \emph{{IEEE} Trans. Veh. Technol.}, vol.~72, no.~5, pp. 6412--6427, 2023.

\bibitem{alishahi2023fl}
M.~Alishahi, P.~Fortier, W.~Hao, X.~Li, and M.~Zeng, ``Energy minimization for wireless-powered federated learning network with {NOMA},'' \emph{{IEEE} Wireless Commun. Lett.}, vol.~12, no.~5, pp. 833--837, 2023.

\bibitem{ren2023ecfl}
Y.~Ren, C.~Wu, and D.~K. So, ``Joint edge association and aggregation frequency for energy-efficient hierarchical federated learning by deep reinforcement learning,'' in \emph{ICC 2023 - IEEE International Commun. Conf. (ICC)}, 2023, pp. 3639--3645.

\bibitem{lin2024cfl}
Y.~Lin, K.~Wang, and Z.~Ding, ``Rethinking clustered federated learning in {NOMA} enhanced wireless networks,'' \emph{arXiv preprint arXiv:2403.03157}, 2024.

\bibitem{rizk2022fl}
E.~Rizk, S.~Vlaski, and A.~H. Sayed, ``Federated learning under importance sampling,'' \emph{{IEEE} Trans. Signal Process.}, vol.~70, pp. 5381--5396, 2022.

\bibitem{yang2020aoi}
H.~H. Yang, A.~Arafa, T.~Q.~S. Quek, and H.~Vincent~Poor, ``Age-based scheduling policy for federated learning in mobile edge networks,'' in \emph{ICASSP 2020 - 2020 IEEE International Conference on Acoustics, Speech and Signal Processing (ICASSP)}, 2020, pp. 8743--8747.

\bibitem{kaidi2022fl}
K.~Wang, Y.~Ma, M.~B. Mashhadi, C.~H. Foh, R.~Tafazolli, and Z.~Ding, ``Age of information in federated learning over wireless networks,'' \emph{arXiv preprint arXiv:2209.06623}, 2022.

\bibitem{boyd2004convex}
S.~Boyd and L.~Vandenberghe, \emph{Convex optimization}.\hskip 1em plus 0.5em minus 0.4em\relax Cambridge university press, 2004.

\end{thebibliography}
%%%%%%%%%%%%%%%%%%%%%%%%%%%%%%%%%%%%%%%%%%%%%%%%%
%%%%%%%%%%%%%%%%%%%%%%%%%%%%%%%%%%%%%%%%%%%%%%%%%
\end{document}